# Revealing non-Hermitian band structures of photonic Floquet media

**Jagang Park[1,†], Hyukjoon Cho[1,†], Seojoo Lee[1†], Kyungmin Lee[1], Kanghee Lee[1], Hee Chul Park[2], Jung-Wan Ryu[2], Namkyoo Park[3], Sanggeun Jeon[4], and Bumki Min[1,*]**

*[1]Department of Mechanical Engineering, Korea Advanced Institute of Science and Technology (KAIST), Daejeon 34141, Republic of Korea*

*[2]Center for Theoretical Physics of Complex Systems, Institute for Basic Science (IBS) Daejeon 34126, Republic of Korea*

*[3]Department of Electrical and Computer Engineering, Seoul National University, Seoul 08826, Republic of Korea*

*[4]School of Electrical Engineering, Korea University, Seoul 02841, Republic of Korea*

[*] Correspondence to: bmin@kaist.ac.kr

[†]These authors contributed equally to this work.

**Periodically driven systems, characterised by their inherent non-equilibrium dynamics, are ubiquitously found in both classical and quantum regimes[1,2]. In the field of photonics, these Floquet systems have begun to provide insight into how time periodicity can extend the concept of spatially periodic photonic crystals and metamaterials to the time domain[3–7]. However, despite the necessity arising from the presence of non-reciprocal coupling between states in a photonic Floquet medium, a unified non-Hermitian band structure description remains elusive. Here, we experimentally reveal the unique Bloch-Floquet and non-Bloch band structures of a photonic Floquet medium emulated in the microwave regime with a one-dimensional array of time-periodically driven resonators. Specifically, these non-Hermitian band structures are shown to be two measurable distinct subsets of complex eigenfrequency surfaces of the photonic Floquet medium defined in complex momentum space. In the Bloch-Floquet band structure, the driving-induced non-reciprocal coupling between oppositely signed frequency states leads to opening of momentum gaps along the real momentum axis, at the edges of which exceptional phase transitions occur[8–10]. More interestingly, we show that the non-Bloch band structure defined in the complex Brillouin zone supplements the information on the morphology of complex eigenfrequency surfaces of the photonic Floquet medium. Our work paves the way for a comprehensive understanding of photonic Floquet media in complex energy-momentum space and could provide general guidelines for the study of non-equilibrium photonic phases of matter.**

Time-periodic driving of a physical system breaks time-reversal and/or time-translational symmetry, thereby enabling observation of exotic new phases of matter in a non- or quasi-equilibrium steady-state setting, such as Floquet quantum Hall states and discrete time crystals[1,2,11]. As in quantum systems ranging from irradiated Dirac matter to semiconductor quantum wells, a spin chain of trapped atomic ions[12] and spin impurities in diamond[13], classical systems can offer a route for observing these phases by virtue of mathematical analogy and the correspondence principle. For example, Floquet quantum Hall states, which were first predicted in irradiated graphene and observed in quantum wells and cold atom arrays[14–16], have been emulated with an array of spatially modulated optical waveguides[17–19] and acoustic metamaterials consisting of dynamically modulated resonators[20–22]. Equally important is that an array of driven coupled non-linear resonators or even a cellular automaton has been recently suggested for observation of discrete time crystalline order in purely classical systems[23,24].

Sophisticatedly designed photonic platforms have been one of the most prevalent and controllable testbeds for demonstration of exotic wave dynamics[25–32] and exhibit tremendous potential for imminent realisation of time-periodically driven effective media. Wave propagation in periodically driven media can be rationalised based on an effective band analysis in the Floquet picture[2]. More specifically, for space-time periodic media, the Bloch-Floquet theorem justifies the existence of Floquet sidebands dressed by the driving. Therefore, the emergence of such dressed states and their interactions in the effective band description are the key to understanding wave propagation in space-time periodic media[3,5,8,33–36]. In view of this, a new perspective on photonic Floquet media is now being provided by advances in non-Hermitian physics. As was theoretically noted, Maxwell's equations with time-periodic non-dispersive permittivity can be recast into an eigenvalue equation with a non-Hermitian Floquet Hamiltonian[8]. More specifically, the driving-induced interaction between two counter-propagating modes is shown to be non-reciprocal, making the reduced Floquet Hamiltonian unitarily equivalent to a parity-time (PT) symmetric Hamiltonian[37,38]. One of the most significant insights provided by non-Hermitian physics is a completely different way of interpreting a momentum gap in a photonic Floquet medium as a broken PT (or unbroken anti-PT) phase with a pair of exceptional points being the gap edges along the real momentum axis[8–10,39]. Nevertheless, the emergence of driving-dressed quasi-frequency bands and resulting interactions in the Bloch-Floquet band structure of photonic Floquet media, which lie at the heart of understanding these emerging photonic platforms, have not been clearly verified, and thorough experimental scrutinisation is yet to come. Here, we take a step further than just experimentally reconstructing the Bloch-Floquet band structure and observing exceptional phase transitions. We show that comprehensive hidden information is embedded in complex eigenfrequency surfaces, the morphology of which can be further revealed by defining and measuring the non-Bloch band structure along a predetermined contour in complex momentum space. These two complementary measurable subsets of complex eigenfrequency surfaces of a photonic Floquet medium could provide a preview of the non-equilibrium dynamics of photonic Floquet media.

**Modelling of photonic Floquet media**

For emulation of a photonic medium with time-periodic permittivity (Fig. 1a), we consider an effective Floquet medium in the microwave regime, in which a waveguide is loaded with a one-dimensional lattice of coupled time-periodically driven $LC$ resonators (Fig. 1b). The necessity of large permittivity modulation leads us to utilise $LC$ resonating unit cells in the construction of the Floquet medium, which, in turn, makes it inherently dispersive. Both the Bloch-Floquet and non-Bloch band structures of this space-time periodic effective medium can be analysed by setting up a time-varying transmission line (TVTL) model consisting of lumped circuit elements. Specifically, by defining nodal variables of the circuit and corresponding modal amplitudes, a Floquet Hamiltonian in momentum space can be derived, from which the exceptional phase transition can be systematically investigated (see Methods and Supplementary Information). For an undriven medium, the hybridisation between a waveguide and a resonator mode results in two static Bloch bands $|\psi_{1+}^{0}\rangle$ and $|\psi_{2+}^{0}\rangle$, drawn with solid black lines in Fig. 1c (also drawn are two mirror symmetric negative frequency bands $|\psi_{1-}^{0}\rangle$ and $|\psi_{2-}^{0}\rangle$, with a line of symmetry being the momentum or wavenumber axis). In the preceding clarification, the static Bloch bands are labelled by $|\psi_{m\pm}^{0}\rangle$, where $m$ denotes the lower ($m = 1$) or higher ($m = 2$) Bloch bands, with the following sign representing either a positive (+) or a negative (-) frequency. As a side note, the validity of the transmission line model is most clearly confirmed by its comparison with the static band structure obtained from full numerical simulations (see Supplementary Figure S1)

**Bloch-Floquet band structures and exceptional phase transitions**

When the time-periodic modulation is turned on, driving-dressed states can exist in the medium; for instance, two such dressed bands, $|\psi_{1-}^{1}\rangle$ and $|\psi_{2-}^{1}\rangle$, are indicated by solid grey lines in Fig. 1c (where the superscript in the band labelling denotes the harmonic order of the driving involved in the formation of the dressed states). As a result, the static undressed states can now be coupled to the dressed states at the band crossing points by a parametric non-reciprocal interaction. In this Bloch-Floquet framework, horizontal momentum gaps are opened due to the non-reciprocal coupling between the undressed state and the driving-dressed state of negative frequency (Fig. 1d). For example, the opening of a primary momentum gap (appearing at half the driving frequency, i.e., $\Omega/2$) is attributed to the non-reciprocal coupling and associated hybridisation between $|\psi_{1+}^{0}\rangle$ and $|\psi_{1-}^{1}\rangle$. Two secondary momentum gaps are symmetrically located along the frequency axis with respect to the primary momentum gap frequency; the upper secondary gap is opened by the hybridisation between $|\psi_{2+}^{0}\rangle$ and $|\psi_{1-}^{1}\rangle$, while the lower secondary gap is opened by the hybridisation between $|\psi_{1+}^{0}\rangle$ and $|\psi_{2-}^{1}\rangle$.

The aforementioned qualitative description is clearly captured by considering the following

2×2 reduced effective Floquet Hamiltonian matrix that describes the interaction between the dressed and undressed states (see Methods and Supplementary Information):

$$H_{red}^{F}(k)=\begin{bmatrix}\omega_{\mu-}(k)+\Omega & \kappa_{\mu\nu}^{F}(k)\\ -\kappa_{\nu\mu}^{F*}(k) & \omega_{\nu+}(k)\end{bmatrix},$$

where $\omega_{\mu-(\nu+)}$ is the modal frequency of the $\mu$-th ($\nu$-th) negative (positive) frequency band, $\Omega$ is the driving frequency, and $\kappa_{\mu\nu}^{F}$ is the driving-induced coefficient of coupling between the $\mu$-th negative and $\nu$-th positive frequency bands. Especially for the interaction between two oppositely signed lower bands $|\psi_{1+}^{0}\rangle$ and $|\psi_{1-}^{1}\rangle$, the reduced Floquet Hamiltonian can be decomposed into the sum of an anti-PT (APT) symmetric matrix and a scalar multiple of the identity matrix,

$$H_{red}^{F}(k)=\frac{\Omega}{2}\mathbf{I}_2+\begin{bmatrix}-\left(\omega_{1+}(k)-\frac{\Omega}{2}\right)^{*} & \kappa_{11}^{F}(k)\\ -\kappa_{11}^{F*}(k) & \omega_{1+}(k)-\frac{\Omega}{2}\end{bmatrix}=\frac{\Omega}{2}\mathbf{I}_2+H_{APT}^{F}(k)$$

Here, the dissipative loss of the resonators was included in the analysis through the use of the Rayleigh dissipation function, which makes the band frequencies $\omega_{\mu-(\nu+)}$ intrinsically complex valued. We can also show that the APT symmetry of the reduced Floquet Hamiltonian matrix is preserved regardless of the inclusion of dissipative loss. Straightforwardly, the eigenvalues (i.e., eigenfrequencies) of the reduced Floquet Hamiltonian matrix are found to be

$$\lambda_{\pm}(k)=\frac{\Omega}{2}+j\mathrm{Im}(\omega_{1+})\pm\sqrt{\left(\mathrm{Re}(\omega_{1+})-\frac{\Omega}{2}\right)^{2}-\mathrm{Im}(\omega_{1+})^{2}-|\kappa_{11}^{F}(k)|^{2}}$$

The eigenfrequency plotted as a function of the real wavenumber signifies two distinct phases of the photonic Floquet medium (Fig. 1e). In the region of real wavenumbers satisfying $|\mathrm{Re}(\omega_{1+})-\Omega/2|<\sqrt{|\mathrm{Im}(\omega_{1+})|^{2}+|\kappa_{11}^{F}|^{2}}$, the two complex eigenfrequencies become conjugate to each other (a broken PT or an unbroken APT phase), while the eigenvalues become distinctive real outside the region (an unbroken PT or a broken APT phase). Accordingly, exceptional phase transitions are explicitly revealed at the two edges of a momentum (or wavenumber) gap, where two mode coalescing singularities along the real momentum axis separate the broken PT region (inside the momentum gap) from two unbroken PT regions (outside the momentum gap). This mode coalescence behaviour of two eigenstates can also be visualised in the Bloch sphere representation (see Fig. 1f). With an increase in the momentum, the eigenstates coalesce and bifurcate at the lower momentum gap edge (corresponding to the south pole in the Bloch sphere), and they coalesce and bifurcate once more at the higher momentum gap edge (i.e., the north pole in the Bloch sphere). Notably, a non-Hermitian Hamiltonian matrix with non-reciprocal and/or dissipative coupling terms in its off-diagonal elements can be unitarily transformed into the matrix with gain and loss terms in its diagonal elements.

**Experimental verification of Bloch-Floquet band structures**

For experimental verification of Bloch-Floquet band structures and exceptional phase transitions, we constructed an effective Floquet medium designed to operate at microwave frequencies by concatenating split-ring resonators inside a customised rectangular waveguide (Fig. 2a). Each resonator is embedded with a DC-biased varactor diode in the capacitive gap; through the application of an AC voltage to the diode terminals, direct modulation of the total capacitance and correspondingly the resonance frequency can be achieved (Fig. 2b, c). Here, the driving strength $\delta_0$ is defined as the modulation depth of the inverse resonator capacitance (see Methods and Supplementary Information). As the first step in verifying the theoretical prediction, Bloch-Floquet band structures were experimentally reconstructed by mapping the spatiotemporal field evolution within the photonic Floquet medium. More specifically, we recorded the temporal evolution of the electric field radiated by a source antenna with a probe antenna scanning over the Floquet medium (see Methods and Supplementary Figure S8). Then, a two-dimensional Fourier transform of the measured electric field was used to reconstruct the Bloch-Floquet band structure. Here, note that the Floquet medium under test was finite sized along the propagation direction; for the first part, the number of unit cells (*N*) was set to 36 to ensure reasonable resolution along the momentum axis while keeping the external driving under optimal conditions. To test the validity of the experimental methodology, we first confirmed the emergence of driving-dressed states $|\psi_{1+}^{1}\rangle$ and $|\psi_{2+}^{1}\rangle$ from the positive frequency lower bands (see Supplementary Figure S2 for the case of driving at 0.6 GHz).

The driving-induced dressing of a negative frequency state, which is of even greater interest, can be resolved in principle once the driving frequency exceeds approximately twice the lower band cutoff, i.e., $\Omega \gtrsim 2\Omega_{c1}$. In Fig. 2d-f, these dressed states of negative frequency and their interaction with undressed states are visualised in the reconstructed Bloch-Floquet band structure, on which theoretically calculated bands are overlaid (the real eigenfrequencies are drawn with semi-transparent violet lines). The opening of momentum gaps can be confirmed by clarifying the locations of high transformed field intensities in the reconstructed Bloch-Floquet band structure, which can be attributed in part to the existence of a temporally growing mode of complex-valued eigenfrequency with a positive imaginary part (see Fig. 1e). This observation can be made clearer by plotting the transformed field intensity as a function of the momentum at half the driving frequency (middle panels of Fig. 2d-f) and comparing it with the calculated imaginary part of the eigenfrequency. According to the comparison, the peak field intensity position on the momentum axis, at which the maximal net gain is expected, is in good agreement with the analysis (bottom panels of Fig. 2d-f). The theoretical prediction is also supported by measured standing-wave-like mode field patterns near the two edges of the primary momentum gap (see Supplementary Figure S3). Specifically, the temporal phase shift of $\sim\pi/2$ in between is analogous to the spatial phase shift of $\pi/2$ between the (energy) band edge modes of a spatially periodic photonic structure.

Here, several more remarks can be made on the momentum gaps in the Bloch-Floquet band structure. First, the primary momentum gap widens as half the driving frequency approaches the frequency of the band edge from the lower side (under the condition of constant driving strength). This is in part due to the modal energy being confined more in the resonators than in the waveguide near the band edges. Second, the approximate mid-gap location of the primary momentum gap lies on the static Bloch band $|\psi_{1+}^{0}\rangle$. This observation can also be confirmed by calculating the mid-gap location as a function of the driving frequency and comparing it with the measured data (Fig. 3a). Finally, the intensity peak at a lower wavenumber ($k \sim 0.06$) observed for a higher driving frequency (the middle and bottom panels of Fig. 2d-f) is indicative of opening of a tertiary gap at the primary gap frequency. This tertiary gap originates from the hybridisation between $|\psi_{2+}^{0}\rangle$ and $|\psi_{2-}^{1}\rangle$ and can be observed for a driving frequency higher than approximately twice the higher band cutoff ($\Omega \gtrsim 2\Omega_{c2}$).

Another compelling argument in support of opening of momentum gaps and occurrence of exceptional transitions is provided by the spectral measurement of noise-initiated in-gap mode parametric oscillation. In these experiments, in contrast to the Bloch-Floquet band structure reconstruction measurements, the power spectral density of the field radiated from the Floquet medium was recorded without any source signal being provided. When the driving strength is increased up to the point where the total loss of the Floquet medium is compensated by the parametric gain within the gap, a noise-initiated oscillating behaviour is induced (see Supplementary Figure S4). Then, the precise frequency of the primary gap and the approximate frequencies of secondary gaps at which the net gain is maximised can be identified by the spectral positions of the parametrically oscillating peaks (see vertical panels in Fig. 2d-f). As predicted, secondary momentum gaps are observed near the spectral positions where $|\psi_{2+}^{0}\rangle$ and $|\psi_{1-}^{1}\rangle$, or $|\psi_{1+}^{0}\rangle$ and $|\psi_{2-}^{1}\rangle$, are presumed to cross in the absence of coupling ($\omega_{1-} + \Omega = \omega_{2+}$, or $\omega_{2-} + \Omega = \omega_{1+}$)[5,35]. By tracking the oscillation frequencies with the variation in the driving frequency (Fig. 3b), we confirmed that the secondary momentum gaps are always symmetrically positioned along the frequency axis with respect to the primary gap, as predicted by theory.

**Complex eigenfrequency surfaces and non-Bloch band structures**

In the aforementioned discussions, all the observed and measured phenomena are interpreted based on the relationship between complex energy and real momentum. Now, we will investigate the characteristics of photonic Floquet media within the non-Bloch framework. For this purpose, we measured the phase retardation of a wave of various frequencies passing through a finite-sized Floquet medium ($N = 24$)[40]. Here, note that this measurement scheme is valid only when the driving strength $\delta_0$ is maintained relatively low; this condition ensures that parametric gain-enhanced multiple reflection can be assumed to be minimal (see Supplementary Figure S5). Within the driving strength regime where the single-pass approximation is valid, a real-valued modal wavenumber can be assigned to the wave of a given (real-valued) frequency as $k_r = \Delta\phi/(Nd)$, where $\Delta\phi$ is the measured phase retardation and $d$ is the thickness of the unit cell. The non-Bloch band structure probed by this measurement protocol enables us to reveal the morphology of complex eigenfrequency surfaces of the Floquet medium in complex momentum space. Notably, the complex eigenfrequency surfaces also contain the Bloch-Floquet band structure as a subset, i.e., the intersections of the surfaces with the plane defined by $\mathrm{Im}(k) = 0$. Therefore, knowledge of both band structures can provide a more in-depth understanding of the photonic Floquet medium. To illustrate this, theoretically calculated real and imaginary eigenfrequency surfaces are plotted in complex momentum space for three different values of the driving strength (Fig. 4a-f). As observed in the plots, the real (or imaginary) eigenfrequency surface is composed of a pair of curved sheets, the characteristics of which are signified by two exceptional points in complex momentum space (see Supplementary Figure S6). In Fig. 4a-f, we can observe that a driving-induced surface morphological change is more pronounced in the region of complex momenta with real values approximately falling in the momentum gap. In contrast, a change in the resonator dissipation rate (see the resistors in Fig. 1b) leads to preferential bending and levelling of the surfaces near the band edge (see Supplementary Figure S6 and S7 for an expanded view of driving- and dissipation-induced changes in the complex eigenfrequency surfaces).

These results prove that the performed phase retardation measurement can be thought of as a selection process that samples only a subset of the real eigenfrequency surfaces in a designated way. The selection rule imposes the condition that the imaginary eigenfrequency must be zero, i.e., $\mathrm{Im}(\omega) = 0$, which can be justified based on the observed steady-state behaviour of the photonic Floquet medium. This steady-state condition enables us to define contours in complex momentum space (as denoted by cyan coloured contours in Fig. 4a-c) and to construct the non-Bloch band structure from the real eigenfrequencies evaluated along these contours (see black lines on the real eigenfrequency surface and their projection onto the sidewalls in Fig. 4d-f). For the case of weak driving ($\delta_0 < \delta_c$), the imaginary part of the eigenfrequency bifurcates within the momentum gap, but the sign is still negative due to dissipation-induced lowering of the eigenfrequency surfaces (see Fig. 4a). Then, the selection rule guarantees phase continuity along the real momentum axis at half the driving frequency (Fig. 4d). In contrast, for the case of strong driving ($\delta_0 > \delta_c$), the sign of a larger bifurcated imaginary eigenfrequency becomes positive near the real momentum axis, changing the zero imaginary eigenfrequency contours such that a phase discontinuity occurs along the real momentum axis at half the driving frequency (see Fig. 4c and f). These two driving regimes are differentiated at a critical driving strength ($\delta_0 = \delta_c$), where the larger of bifurcated imaginary eigenfrequencies becomes zero (Fig. 4b and e).

**Complex Brillouin zone**

The difference between non-Bloch and Bloch-Floquet band structures can be clearly noted by plotting the corresponding Bloch phase factors $\beta \equiv \exp(jk)$ for two subsets of complex momentum defined by $\mathrm{Im}(\omega) = 0$ and $\mathrm{Im}(k) = 0$. The latter condition ensures that the Bloch phase factor $\beta_{\mathrm{Im}(k)=0}$ takes continuous values on a unit circle centred at the origin of the Gauss plane (drawn with black dashed lines in Fig. 5a-c), which represents the conventional Brillouin zone for the Bloch-Floquet band structure. In contrast, the Bloch phase factor $\beta_{\mathrm{Im}(\omega)=0}$ does not trace the unit circle (drawn with cyan lines in Fig. 5a-c), the consequences of which are reflected in the non-Bloch band structure measured in the steady state. More specifically, for the case of strong driving ($\delta_0 > \delta_c$), the trace of $\beta_{\mathrm{Im}(\omega)=0}$ no longer encircles the origin but exhibits gaps in the angular direction (highlighted by yellow sectors in Fig. 5c). The angular gaps in the trace imply that there exists a finite range of $\mathrm{Re}(k)$ for which real-valued eigenfrequencies are absent. Therefore, the non-Bloch band structure projected onto the $\mathrm{Re}(k)$-$\mathrm{Re}(\omega)$ plane exhibits a discontinuous jump along the real momentum axis at the primary momentum gap frequency. This is in stark contrast to the immediate opening of momentum gaps in the Bloch-Floquet band structure, which is observed regardless of the driving strength (see Fig. 4d-i). As demonstrated, the non-Bloch band structure reconstruction, when combined with variations in parameters (e.g., driving strength or the rate of dissipation), can provide sectional information on the morphology of complex eigenfrequency surfaces in complex momentum space, which calls for a further in-depth investigation on how the non-Hermiticity of photonic Floquet media manifests itself in complex momentum space[41–43].

Although one of the most intriguing features of a photonic Floquet medium is the creation of dressed states of negative frequency and their interactions leading to exceptional phase transitions, importantly, the interpretation of experimental observations given here is based on a linearised model. In the strong driving regime, however, ignoring non-linearity along with dissipation and fluctuation becomes increasingly difficult. This implies that a thorough understanding of these effects is required for characterisation of exotic new phases in non-equilibrium photonic matter. In this regard, non-reciprocal phase transitions in active matter seem to share many features in common with the proposed non-equilibrium photonic Floquet medium, which coherently exchanges energy with its environment[24,44,45]. Particularly, classical many-body effects can be manifested by non-linearity; whether a subharmonic spectral component arising at the momentum gap is associated with many-body time crystalline behaviour would be an especially interesting question for which further investigation is highly demanded. Considered from another perspective, the proposed platform is reminiscent of classical coherent Ising machines constructed by coupling non-linear oscillators for solving computationally hard problems[46]. The complex band structure analysis also provides a new perspective on designing driven nonlinear photonic devices such as parametric amplifiers, oscillators and isolators. Finally, photonic Floquet media are an effective spatiotemporal matter platform generalised from conventional spatially periodic structures and are poised to create a new area of research in non-Hermitian photonics.

## References

1. Basov, D. N., Averitt, R. D. & Hsieh, D. Towards properties on demand in quantum materials. *Nat. Mater.* **16**, 1077–1088 (2017).

2. Oka, T. & Kitamura, S. Floquet engineering of quantum materials. *Annu. Rev. Condens. Matter Phys.* **10**, 387–408 (2019).

3. Martínez-Romero, J. S., Becerra-Fuentes, O. M. & Halevi, P. Temporal photonic crystals with modulations of both permittivity and permeability. *Phys. Rev. A* **93**, (2016).

4. Caloz, C. & Deck-Leger, Z. L. Spacetime metamaterials - Part I: General concepts. *IEEE Trans. Antennas Propag.* **68**, 1569–1582 (2020).

5. Chamanara, N., Deck-Léger, Z. L., Caloz, C. & Kalluri, D. Unusual electromagnetic modes in space-time-modulated dispersion-engineered media. *Phys. Rev. A* **97**, (2018).

6. Lustig, E., Sharabi, Y. & Segev, M. Topological aspects of photonic time crystals. *Optica* **5**, 1390–1395 (2018).

7. Lee, K. *et al.* Linear frequency conversion via sudden merging of meta-atoms in time-variant metasurfaces. *Nat. Photonics* **12**, 765–773 (2018).

8. Wang, N., Zhang, Z. Q. & Chan, C. T. Photonic Floquet media with a complex time-periodic permittivity. *Phys. Rev. B* **98**, (2018).

9. Koutserimpas, T. T. & Fleury, R. Electromagnetic fields in a time-varying medium: Exceptional points and operator symmetries. *IEEE Trans. Antennas Propag.* **68**, 6717–6724 (2020).

10. Rouhi, K., Kazemi, H., Figotin, A. & Capolino, F. Exceptional points of degeneracy directly induced by space-time modulation of a single transmission line. *IEEE Antennas Wirel. Propag. Lett.* **19**, 1906–1910 (2020).

11. Else, D. V., Monroe, C., Nayak, C. & Yao, N. Y. Discrete time crystals. *Annu. Rev. Condens. Matter Phys.* **11**, 467–499 (2020).

12. Zhang, J. *et al.* Observation of a discrete time crystal. *Nature* **543**, 217–220 (2017).

13. Choi, S. *et al.* Observation of discrete time-crystalline order in a disordered dipolar many-body system. *Nature* **543**, 221–225 (2017).

14. Oka, T. & Aoki, H. Photovoltaic Hall effect in graphene. *Phys. Rev. B* **79**, 081406 (2009).

15. Lindner, N. H., Refael, G. & Galitski, V. Floquet topological insulator in semiconductor quantum wells. *Nat. Phys.* **7**, 490–495 (2011).

16. Wintersperger, K. *et al.* Realization of an anomalous Floquet topological system with ultracold atoms. *Nat. Phys.* **16**, 1058–1063 (2020).

17. Rechtsman, M. C. *et al.* Photonic Floquet topological insulators. *Nature* **496**, 196–200 (2013).

18. Maczewsky, L. J., Zeuner, J. M., Nolte, S. & Szameit, A. Observation of photonic anomalous Floquet topological insulators. *Nat. Commun.* **8**, 13756 (2017).

19. He, L. *et al.* Floquet Chern insulators of light. *Nat. Commun.* **10**, (2019).

20. Fleury, R., Khanikaev, A. B. & Alù, A. Floquet topological insulators for sound. *Nat. Commun.* **7**, (2016).

21. Chen, H., Yao, L. Y., Nassar, H. & Huang, G. L. Mechanical Quantum Hall Effect in Time-Modulated Elastic Materials. *Phys. Rev. Appl.* **11**, 044029 (2019).

22. Xia, Y. *et al.* Experimental Observation of Temporal Pumping in Electromechanical Waveguides. *Phys. Rev. Lett.* **126**, (2021).

23. Yao, N. Y., Nayak, C., Balents, L. & Zaletel, M. P. Classical discrete time crystals. *Nat. Phys.* **16**, 438–447 (2020).

24. Fruchart, M., Hanai, R., Littlewood, P. B. & Vitelli, V. Non-reciprocal phase transitions. *Nature* **592**, 363–369 (2021).

25. Ozawa, T., Price, H. M., Goldman, N., Zilberberg, O. & Carusotto, I. Synthetic dimensions in integrated photonics: From optical isolation to four-dimensional quantum Hall physics. *Phys. Rev. A* **93**, 1–17 (2016).

26. Yuan, L., Lin, Q., Xiao, M. & Fan, S. Synthetic dimension in photonics. *Optica* **5**, 1396 (2018).

27. Shaltout, A. M., Shalaev, V. M. & Brongersma, M. L. Spatiotemporal light control with active metasurfaces. *Science (80-. ).* **364**, eaat3100 (2019).

28. Zhou, Y. *et al.* Broadband frequency translation through time refraction in an epsilon-near-zero material. *Nat. Commun.* **11**, 2180 (2020).

29. Mukherjee, S. & Rechtsman, M. C. Observation of Floquet solitons in a topological bandgap. *Science (80-. ).* **368**, 856–859 (2020).

30. Sounas, D. L. & Alù, A. Non-reciprocal photonics based on time modulation. *Nat. Photonics* **11**, 774–783 (2017).

31. Wang, K. *et al.* Generating arbitrary topological windings of a non-Hermitian band. *Science (80-. ).* **371**, 1240–1245 (2021).

32. Wang, K., Dutt, A., Wojcik, C. C. & Fan, S. Topological complex-energy braiding of non-Hermitian bands. *Nature* **598**, 59 (2021).

33. Wang, Y. H., Steinberg, H., Jarillo-Herrero, P. & Gedik, N. Observation of Floquet-Bloch states on the surface of a topological insulator. *Science (80-. ).* **342**, 453–457 (2013).

34. Riva, E., Marconi, J., Cazzulani, G. & Braghin, F. Generalized plane wave expansion method for non-reciprocal discretely modulated waveguides. *J. Sound Vib.* **449**, 172–181 (2019).

35. Lee, S. *et al.* Parametric oscillation of electromagnetic waves in momentum band gaps of a spatiotemporal crystal. *Photonics Res.* **9**, 142 (2021).

36. Park, J. & Min, B. Spatiotemporal plane wave expansion method for arbitrary space–time periodic photonic media. *Opt. Lett.* **46**, 484–487 (2021).

37. Zhang, F., Feng, Y., Chen, X., Ge, L. & Wan, W. Synthetic anti-PT Symmetry in a single microcavity. *Phys. Rev. Lett.* **124**, 1–6 (2020).

38. Bergman, A. *et al.* Observation of anti-parity-time-symmetry, phase transitions and exceptional points in an optical fibre. *Nat. Commun.* **12**, (2021).

39. Kazemi, H., Nada, M. Y., Mealy, T., Abdelshafy, A. F. & Capolino, F. Exceptional points of degeneracy induced by linear time-periodic variation. *Phys. Rev. Appl.* **11**, 014007 (2019).

40. Reyes-Ayona, J. R. & Halevi, P. Observation of genuine wave vector (k or β) gap in a dynamic transmission line and temporal photonic crystals. *Appl. Phys. Lett.* **107**, (2015).

41. Yokomizo, K. & Murakami, S. Non-Bloch band theory of non-Hermitian systems. *Phys. Rev. Lett.* **123**, 66404 (2019).

42. Kawabata, K., Okuma, N. & Sato, M. Non-Bloch band theory of non-Hermitian Hamiltonians

in the symplectic class. *Phys. Rev. B* **101**, 195147 (2020).

43. Scheibner, C., Irvine, W. T. M. & Vitelli, V. Non-Hermitian band topology and skin modes in active elastic media. *Phys. Rev. Lett.* **125**, 118001 (2020).

44. Sone, K., Ashida, Y. & Sagawa, T. Exceptional non-Hermitian topological edge mode and its application to active matter. *Nat. Commun.* **11**, 1–11 (2020).

45. Pendry, J. B., Galiffi, E. & Huidobro, P. A. Gain mechanism in time-dependent media. *Optica* **8**, 636–637 (2021).

46. Marandi, A., Wang, Z., Takata, K., Byer, R. L. & Yamamoto, Y. Network of time-multiplexed optical parametric oscillators as a coherent Ising machine. *Nat. Photonics* **8**, 937–942 (2014).

## Methods

### Transmission line modelling of static band structures

We established a TVTL model with lumped circuit elements to analyse Floquet media, the properties of which are simultaneously dispersive and time varying. The static band structure can be obtained simply by assuming that none of the constituting circuit elements of the TVTL are time varying. The model consists of a chain of driven $LC$ resonators and a bus waveguide, as depicted in Fig. 1b. The constituting resonators are assumed to be capacitively coupled to the waveguide and inductively coupled to their nearest neighbours. The Lagrangian can be given as a function of accumulated charge and electric current ($Q$ and $\dot{Q}$) at every nodal point and circuit element of the TVTL:

$$\mathcal{L} = \sum_{n} \left( \frac{1}{2} L_0 \left| \dot{Q}_n^{L_0} \right|^2 + \frac{1}{2} L_s \left| \dot{Q}_n^{L_s} \right|^2 + \frac{1}{2} L_r \left| \dot{Q}_n^{L_r} \right|^2 + \sum_{i} M_i \dot{Q}_n^{L_r} \dot{Q}_{n+i}^{L_r} - \frac{|Q_n^a|^2}{2C_0} - \frac{|Q_n^b|^2}{2C_r} - \frac{Q_n^a Q_n^b}{C_g} \right)$$

where $L_0$, $L_s$, $C_0$, $L_r$, $C_r$, $M_i$, and $C_g$ are the series inductance, shunt inductance and shunt capacitance of the bus waveguide, inductance and capacitance of the resonator, mutual inductance between $i$-th nearest-neighbour resonators and coupling capacitance between the resonator and the bus waveguide. The superscripts on $\dot{Q}$s indicate the circuit elements conducting the electric current, while the superscripts $a$ and $b$ on $Q$s denote the waveguide and resonator nodes, respectively. The Legendre transformation of the Fourier transformed Lagrangian gives the Hamiltonian in momentum space as a function of charge and node flux ($Q$ and $\Phi$), which are the canonical variables for electric circuits. By defining modal amplitudes of the waveguide ($a_\kappa$) and the coupled resonators ($b_\kappa$) as linear combinations of these canonical variables, the Hamiltonian can be written as

$$H_\kappa = \Omega_\kappa \left( a_\kappa^\dagger a_\kappa + a_{-\kappa}^\dagger a_{-\kappa} \right)/2 + \omega_\kappa \left( b_\kappa^\dagger b_\kappa + b_{-\kappa}^\dagger b_{-\kappa} \right)/2 - g_\kappa \left[ \left( a_\kappa^\dagger + a_{-\kappa} \right)\left( b_\kappa + b_{-\kappa}^\dagger \right) + \left( b_\kappa^\dagger + b_{-\kappa} \right)\left( a_\kappa + a_{-\kappa}^\dagger \right) \right]/2,$$

where $\Omega_\kappa$ and $\omega_\kappa$ are the frequencies of the bus waveguide and the coupled resonator with momentum $\kappa$, respectively, and $g_\kappa$ is the coefficient of coupling between these two modes, which are found as follows:

$$\Omega_\kappa = \sqrt{1/L_s C_0 + [4\sin^2(\pi/N)\kappa]/L_0 C_0},$$

$$\omega_\kappa = 1/\sqrt{[L_r + \Sigma_i M_i \cos(2\pi/N)\kappa i]C_r},$$

$$g_\kappa = \sqrt{C_0 C_r \Omega_\kappa \omega_\kappa}/2C_g.$$

On the basis of mode amplitudes $(a_\kappa, b_\kappa, a_{-\kappa}^\dagger, b_\kappa^\dagger)$, the effective Hamiltonian is found in the form of a matrix as

$$\widetilde{H}_\kappa = \frac{1}{2}\begin{bmatrix} \Omega_\kappa & -g_\kappa & 0 & -g_\kappa \\ -g_\kappa & \omega_\kappa & -g_\kappa & 0 \\ 0 & g_\kappa & -\Omega_\kappa & g_\kappa \\ g_\kappa & 0 & g_\kappa & -\omega_\kappa \end{bmatrix}.$$

Then, by calculating eigenvalues of $\widetilde{H}_\kappa$, the exact static band structure of the model can be found to be

$$\omega_1^2 = \frac{1}{2}\left[(\Omega_\kappa^2 + \omega_\kappa^2) - \sqrt{(\Omega_\kappa^2 - \omega_\kappa^2)^2 + 16\Omega_\kappa \omega_\kappa g_\kappa^2}\right]$$

$$\omega_2^2 = \frac{1}{2}\left[(\Omega_\kappa^2 + \omega_\kappa^2) + \sqrt{(\Omega_\kappa^2 - \omega_\kappa^2)^2 + 16\Omega_\kappa \omega_\kappa g_\kappa^2}\right]$$

For the sake of simplicity, an approximated Hamiltonian can be obtained by applying the rotating-wave approximation:

$$H_\kappa = \Omega_\kappa a_\kappa^\dagger a_\kappa + \omega_\kappa b_\kappa^\dagger b_\kappa - g_\kappa\left(b_\kappa^\dagger a_\kappa + a_\kappa^\dagger b_\kappa\right),$$

which can be used to derive approximate static band structures of the transmission line model:

$$\omega_{1\pm} = \pm\frac{1}{2}\left[(\Omega_\kappa + \omega_\kappa) - \sqrt{(\Omega_\kappa - \omega_\kappa)^2 + 4g_\kappa^2}\right] \approx \pm\sqrt{\omega_1^2}$$

$$\omega_{2\pm} = \pm\frac{1}{2}\left[(\Omega_\kappa + \omega_\kappa) + \sqrt{(\Omega_\kappa - \omega_\kappa)^2 + 4g_\kappa^2}\right] \approx \pm\sqrt{\omega_2^2}$$

These approximated static band structures are drawn in Fig. 1c.

**Bloch-Floquet band structures of the TVTL**

For the calculation of Bloch-Floquet band structures, we first assumed that the resonator capacitance is periodically driven as $C_r(t) = C_c/[1+\delta(t)]$ with a temporal periodicity of $2\pi/\Omega$; this assumption makes the Hamiltonian time periodic, i.e., $H_\kappa(t+2\pi/\Omega) = H_\kappa(t)$. The temporal periodicity allows us to use the Fourier series expansion of the Hamiltonian, $H_\kappa(t) = \sum_q \exp(-jq\Omega t)\, H_\kappa^q$ , and choose the solution of the Floquet form $\psi_\kappa(t) = \exp(-j\omega t)\sum_q \exp(-jq\Omega t)\psi_\kappa^q$. Consequently, the Hamilton's equations can be rearranged into a set of time-independent equations using the orthogonality of harmonic terms. The equations can be recast into an eigenvalue problem with the following time-independent effective Floquet Hamiltonian matrix:

$$\widetilde{H}_\kappa^F = \begin{bmatrix} \ddots & & & & \\ & \widetilde{H}_\kappa^0 + \Omega \mathrm{I} & \widetilde{H}_\kappa^{-1} & \widetilde{H}_\kappa^{-2} & \\ & \widetilde{H}_\kappa^{+1} & \widetilde{H}_\kappa^0 & \widetilde{H}_\kappa^{-1} & \\ & \widetilde{H}_\kappa^{+2} & \widetilde{H}_\kappa^{+1} & \widetilde{H}_\kappa^0 - \Omega \mathrm{I} & \\ & & & & \ddots \end{bmatrix}$$

The size of the Floquet Hamiltonian matrix depends on the harmonic order considered in the analysis and determines the number of shifted replicas of static bands appearing in the band structure. In the case where $\delta(t)$ is sinusoidal, i.e., $\delta(t) = \delta_0 \cos(\Omega t + \varphi_0)$, we can truncate the Floquet Hamiltonian matrix up to the first harmonic order without significant loss of accuracy. The calculated Bloch-Floquet band structures are shown in Fig. 1d and overlaid on the experimentally measured band structure in Fig. 2.

**Reduced Floquet Hamiltonian**

The opening of a gap along the momentum axis is attributed to the non-reciprocal coupling between two dominant bands participating in the interaction, one of which is the undressed (static) band and the other is the dressed band. Therefore, the driving-induced interaction between the two bands can be intuitively understood by constructing a 2×2 effective

Hamiltonian matrix, the so-called reduced Floquet Hamiltonian[8]. With the rotating-wave approximation followed by a basis transformation, the static component of the Floquet Hamiltonian can be diagonalised; $\bar{H}_\kappa^0 = P^{-1}\tilde{H}_\kappa^0 P = \mathrm{diag}(\omega_{i\pm})$. Then, the reduced Floquet Hamiltonian is obtained by choosing only four elements from the entire Floquet Hamiltonian considering the driving-induced interaction between the involved bands. For the lossless case, the non-reciprocal coupling coefficients for the dressed $\mu$-th band and the undressed $\nu$-th band are found to be

$$\kappa_{\mu\nu}^F = \frac{\Delta_\nu}{\Delta_2^* - \Delta_1^*}\frac{\omega_\kappa \delta_0}{4} e^{j\varphi_M}$$

where $\Delta_{\mu(\nu)} \equiv \Omega_\kappa - \omega_{\mu+(\nu+)}$. The reduced Bloch-Floquet band structures calculated from the reduced Floquet Hamiltonian are used for the non-Hermitian physics-based analyses in the main manuscript.

**Inclusion of dissipative loss in resonators**

The resonator dissipative loss can be considered by introducing the Rayleigh dissipation function and adding the corresponding non-conservative force into the equation of motion. A serially connected resistor is included in each constituting *LC* resonator as a source of dissipation. Then, the Rayleigh dissipation function in momentum space is found to be

$$\mathcal{F}_\kappa = \frac{1}{2}R\left|\dot{Q}_\kappa^b\right|^2 = \gamma_\kappa \omega_\kappa \left(b_\kappa^\dagger - b_{-\kappa}\right)\left(b_{-\kappa}^\dagger - b_\kappa\right)$$

where $\gamma_\kappa$ is the loss rate proportional to the resistance. By including the dissipative loss, the frequency of the static band becomes intrinsically complex valued as follows:

$$\omega_{1+} = \frac{1}{2}\left[(\Omega_\kappa + \omega_\kappa - j\gamma_\kappa) - \sqrt{(\Omega_\kappa - \omega_\kappa + j\gamma_\kappa)^2 + 4g_\kappa^2}\right] = -\omega_{1-}^*$$

$$\omega_{2+} = \frac{1}{2}\left[(\Omega_\kappa + \omega_\kappa - j\gamma_\kappa) + \sqrt{(\Omega_\kappa - \omega_\kappa + j\gamma_\kappa)^2 + 4g_\kappa^2}\right] = -\omega_{2-}^*$$

The band structures considering the dissipative loss were used to interpret the phase retardation measurements and non-Bloch band structures shown in Figs. 4 and 5.

## Fabrication of samples and preparation of the measurement setup

The split-ring resonators comprising the photonic Floquet medium were patterned on low-loss dielectric substrates (RO4350B 30MIL, Rogers) by a printed circuit board (PCB) manufacturing process. Then, a varactor diode (SMV-1247, Skyworks) was soldered to the resonator split gap and connected via an SMA connector to external driving circuitry for periodic capacitance modulation (see Supplementary Figure S2 for the design of unit cells and their operation). The driving signal was generated and amplified by a signal generator (E4438C, Keysight) and an RF amplifier (RUM43020-10, RFHIC), while its power was monitored before distributing the signal to the constituting unit cells. Each resonator was DC biased using a bias tee (ZX85-40W-63-S+, Mini-circuits), and the operating DC bias voltage was optimised to ensure large opening of a primary momentum gap. For spatiotemporal electric field probe scanning measurements, a rectangular waveguide with a DC servo-controlled movable upper plate was custom-made, and a quarter-wave monopole antenna was attached to the movable upper plate. In addition, a commercially available waveguide was used for the phase retardation and parametric oscillation measurements. The cutoff frequency of both waveguides was 1.372 GHz.

## Bloch-Floquet band structure construction by spatiotemporal field probing

The number of unit cells used to probe the Bloch-Floquet band structure was set to 36 to ensure maximal momentum space resolution under realistic driving conditions. In all the measurements, the distance between the nearest-neighbour unit cells was fixed to 10 mm. To provide a seeding field for spatiotemporal modal evolution inside the Floquet medium, an on-off keying signal (N5171B, Keysight) with a variable (angular) carrier frequency ($\omega$) was fed into the input port of the waveguide (see Supplementary Figure S8). To synchronise the acquisition at different spatial probe positions, a rising signal edge was used to trigger the whole measurement sequence. Experimental data were monitored with a spectrum analyser (E4404B, Keysight) and a digital oscilloscope (DPO70804, Tektronix). To obtain high resolution along the frequency axis, the input frequency was scanned in 10 MHz steps. Then, the acquired spatiotemporal field was Fourier transformed to reconstruct the Bloch-Floquet band structures. For clear visualisation of dressed and undressed states, we mainly utilised three spatially resolved spectral components ($\omega, \Omega - \omega$ and $\Omega/2$), and the Bloch-Floquet band was reconstructed by merging each spectral component with normalised amplitude on the frequency axis.

**Non-Bloch band construction by phase retardation measurement**

For characterisation of phase retardation and construction of non-Bloch band structures, three measurements were sequentially performed (see Supplementary Figure S8). First, the phase retardation through an empty waveguide and a static resonator-loaded waveguide was measured as a function of frequency. Here, the measurement on the empty waveguide was used to subtract the phase retardation accumulated along the path outside the Floquet medium (i.e., the path from the source to the medium and the medium to a detector). We first checked the validity of this measurement scheme by reconstructing the static band structure and comparing it with the theoretically calculated structure. With driving on, the same measurement gives the non-Bloch band structures shown in the main manuscript (Fig. 5). For clear visualisation of the driving-induced dressing of negative frequency states, we mirror imaged the measured curves with respect to the momentum gap frequency.

**Acknowledgements**

We thank Prof. Sunkyu Yu, Dr. Xianji Piao, and Dr. Joonhee Choi for helpful discussions. This work was supported by National Research Foundation of Korea (NRF) through the government of Korea (NRF-2017R1A2B3012364). The work was also supported by the center for Advanced Meta-Materials (CAMM) funded by Korea Government (MSIP) as Global Frontier Project (NRF-2014M3A6B3063709).

**Author contributions**

B.M. conceived the original idea and supervised the project. J.P. carried out the theoretical analysis. H.C. and S.L. designed the unit cell through numerical calculations. H.C., S.L., and K.M.L. performed measurements. K.H.L. and S.J. helped and guided the measurements. All authors discussed the theoretical and experimental results. J.P. and B.M. wrote the manuscript, and all authors provided feedback.

**Competing financial interests**

The authors declare no competing financial interests.

**Figure captions**

**Figure 1 Bloch-Floquet band structures and exceptional transitions.** (a) Homogeneous Floquet medium with time-periodic permittivity. (b) Schematic illustration of a one-dimensional effective photonic Floquet medium consisting of spatially concatenated time-varying *LC* resonators and a waveguide. The resonator capacitance and correspondingly the resonance frequency are assumed to be time periodically modulated. Each of the resonators is capacitively coupled to the waveguide and inductively coupled to its nearest-neighbour resonators. (c) Static band structure (drawn with black lines) of the undriven medium and creation of driving-dressed states (drawn with grey lines) of negatively signed frequency in the photonic Floquet medium. The resonator and waveguide bands are drawn with dashed-dotted and dashed lines, respectively. (d) Formation of the primary (drawn with green lines) and secondary (drawn with blue and red lines) momentum gaps resulting from the interaction between undressed and dressed states. Dashed lines are calculated without considering the interactions. (e) Complex eigenfrequencies plotted as a function of the real wavenumber in the proximity of the primary momentum gap. Mode coalescence behaviour and exceptional transitions are observed at the edges of the primary momentum gap. The momentum gap is associated with the APT unbroken region. (f) Bloch sphere representation of eigenmodes $|\lambda_{\pm}\rangle$ at various wavenumbers denoted by scatters in (e). The mode coalescence and bifurcation occur at the north and south poles of the Bloch sphere.

**Figure 2 Experimental reconstruction of the Bloch-Floquet band structure.** (a) Schematic illustration of the microwave Floquet medium comprising an array of driven *LC* resonators in a rectangular waveguide. (b) Geometry of a split-ring *LC* resonator patterned by a PCB manufacturing process on a low-loss dielectric substrate. A varactor diode is soldered to the capacitive gap of the resonator. The geometrical parameters of the unit cell are designed as follows: $a_x$ = 80 mm, $a_y$ = 30 mm, $w$ = 10 mm, and $g$ = 1 mm. The spacing between the unit cells is set to 10 mm. Below is the plot of the inverse capacitance as a function of reverse bias voltage applied to the varactor diode (SMV1247, Skyworks). In the measurements, the Q-point was set by maximising the signal parametrically oscillating at the primary momentum gap frequency. The orange line represents the variation in the inverse capacitance for the sinusoidal driving voltage excitation drawn with a green line. (c) Averaged transmission spectra of the unit cells for bias voltages of 1 V, 3 V and 5 V. Grey lines show individually measured transmission spectra of 24 different unit cells. (d-f) Two-dimensional Fourier transformed electric fields for modulation frequencies of (d) 4.6 GHz, (e) 4.65 GHz and (f) 4.7 GHz. Theoretically calculated band structures are overlaid on the measured results with semi-transparent violet lines. Middle panels are the horizontal cut of the transformed field along the real momentum axis at the primary momentum gap frequency. Lower panels are the imaginary part of the eigenfrequency plotted as a function of wavenumber. The power spectral density of the radiation from the strongly driven Floquet medium is shown in the right panel. Sharp oscillating peaks are observed at momentum gap frequencies.

**Figure 3 Identification of momentum gap locations in the Bloch-Floquet band structure.** (a) Plot of mid-gap wavenumbers as a function of the driving frequency (for the primary momentum gap). Theoretically calculated mid-gap wavenumbers at which the imaginary part of the complex eigenfrequency becomes maximised are drawn with a black line. Experimentally measured mid-gap wavenumbers are estimated from the locations of Fourier transformed intensity maxima at the primary momentum gap frequencies. (b) Plot of momentum gap frequencies as a function of the driving frequency. Primary and secondary

momentum gap frequencies are estimated from the experimentally measured parametric oscillation frequencies (see Fig. 2d-f) and compared with the theoretical predictions from the TVTL model (drawn with lines). Parametric oscillations are attributed to the loss-compensated unstable modes inside the primary (green) and secondary (upper; blue, lower; red) momentum gaps.

**Figure 4 Complex eigenfrequency surfaces and non-Bloch band structures.** (a-c) Imaginary eigenfrequency surfaces plotted in complex momentum space for three distinct values of the driving strength, i.e., weak driving ($\delta_0 < \delta_c$), critical driving ($\delta_0 = \delta_c$), and strong driving ($\delta_0 > \delta_c$). The imaginary eigenfrequency surfaces consisting of a pair of curved sheets are colour coded by the values of their imaginary part. The zero imaginary eigenfrequency contours are drawn with cyan coloured lines. (d-f) Real eigenfrequency surfaces plotted in complex momentum space for three distinct values of the driving strength, i.e., weak driving ($\delta_0 < \delta_c$), critical driving ($\delta_0 = \delta_c$), and strong driving ($\delta_0 > \delta_c$). The real eigenfrequency surfaces consisting of a pair of curved sheets are colour coded by the values of their corresponding imaginary eigenfrequency. Here, the black lines (drawn on the surface) are the real eigenfrequencies corresponding to the complex momenta on cyan coloured contours (i.e., zero imaginary eigenfrequency contours). Shown on the sidewall are the non-Bloch band structures that can be obtained from phase retardation measurements (see Fig. 5). Note that phase discontinuity is observed only for strong driving (highlighted by a yellow-coloured area).

**Figure 5 Comparison between non-Bloch and Bloch-Floquet band structures.** (a-c) Traces of the Bloch phase factor $\beta = \exp(jk)$ plotted on the complex plane for two different subsets

of complex momentum. One is based on the conventional definition of the Brillouin zone (denoted by a dashed black unit circle; $|\beta| = 1$), and the other is based on the extended Brillouin zone from steady-state phase retardation measurements (drawn with a cyan solid line; $\mathrm{Im}(\omega) = 0$). Here, the complex Brillouin zone is introduced for interpretation of measured phase retardation data. For strong driving, angular gaps in the traces of the Bloch phase factor are highlighted with yellow sectors. (d-f) Theoretically calculated Bloch-Floquet band structure defined on the real momentum axis (denoted by dashed black lines) and non-Bloch band structures (denoted by cyan solid lines). Note that the momentum gap appears in the Bloch-Floquet band structure regardless of the driving strength. The non-Bloch band structures are reconstructed from the phase retardation measurements at three different values of the driving strength approximately corresponding to (d) weak, (e) critical, and (f) strong driving (with the power being measured before splitting the driving signal). Theoretically predicted non-Bloch band structures are well reproduced in the experiments. The yellow region in (f) is associated with the angular gaps shown in (c).

## Figures

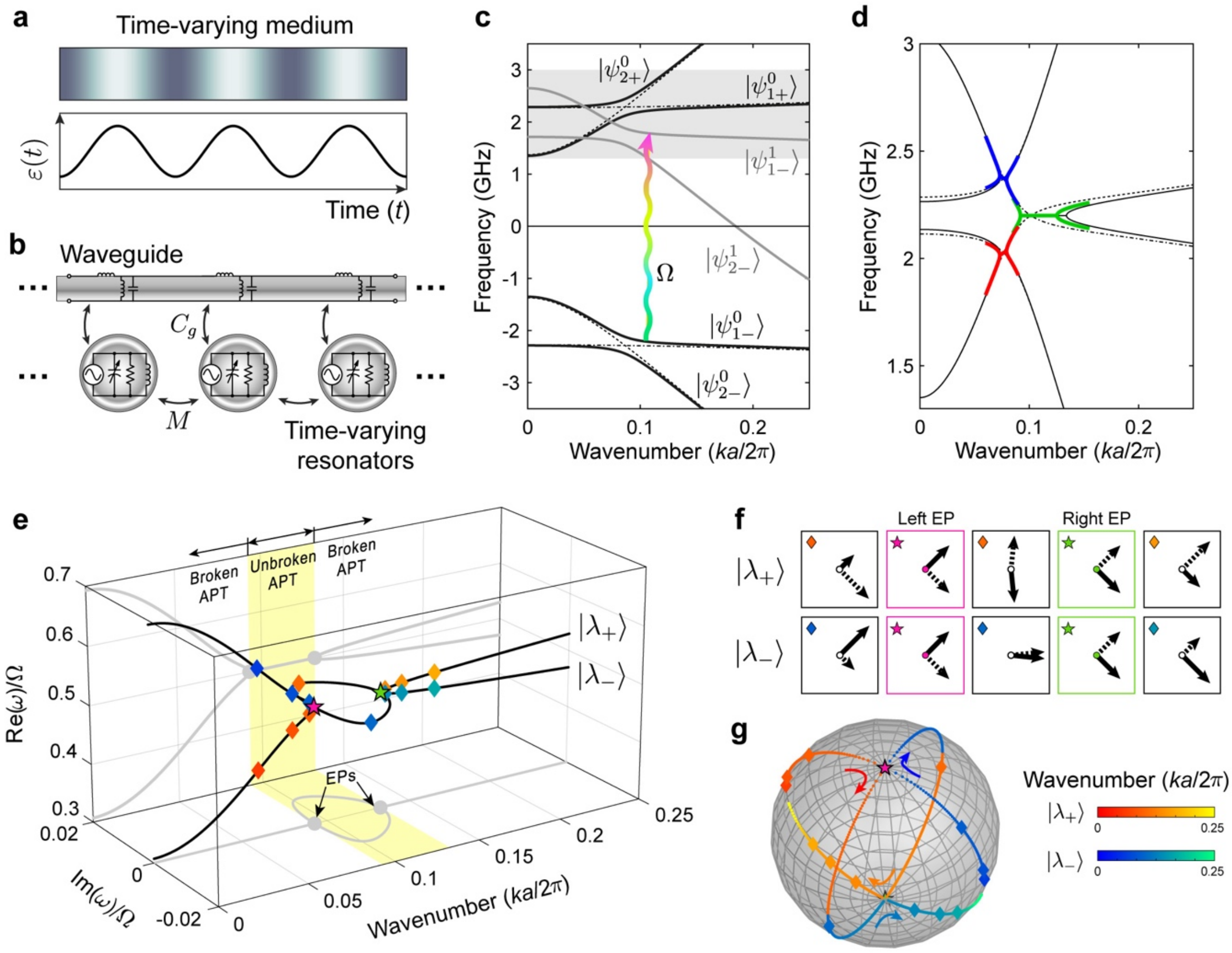

Fig. 1 Park et al.

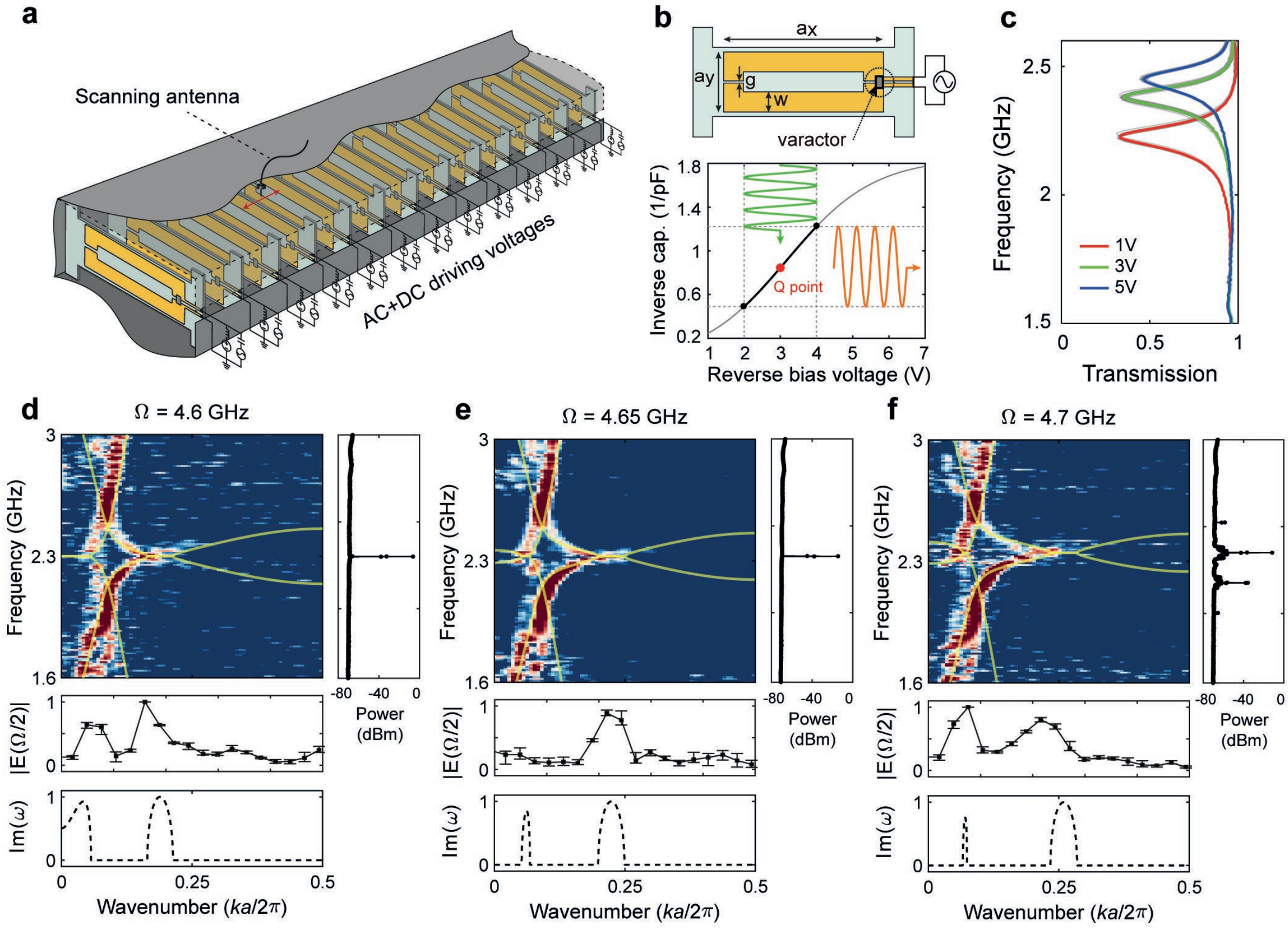

Fig. 2 Park et al.

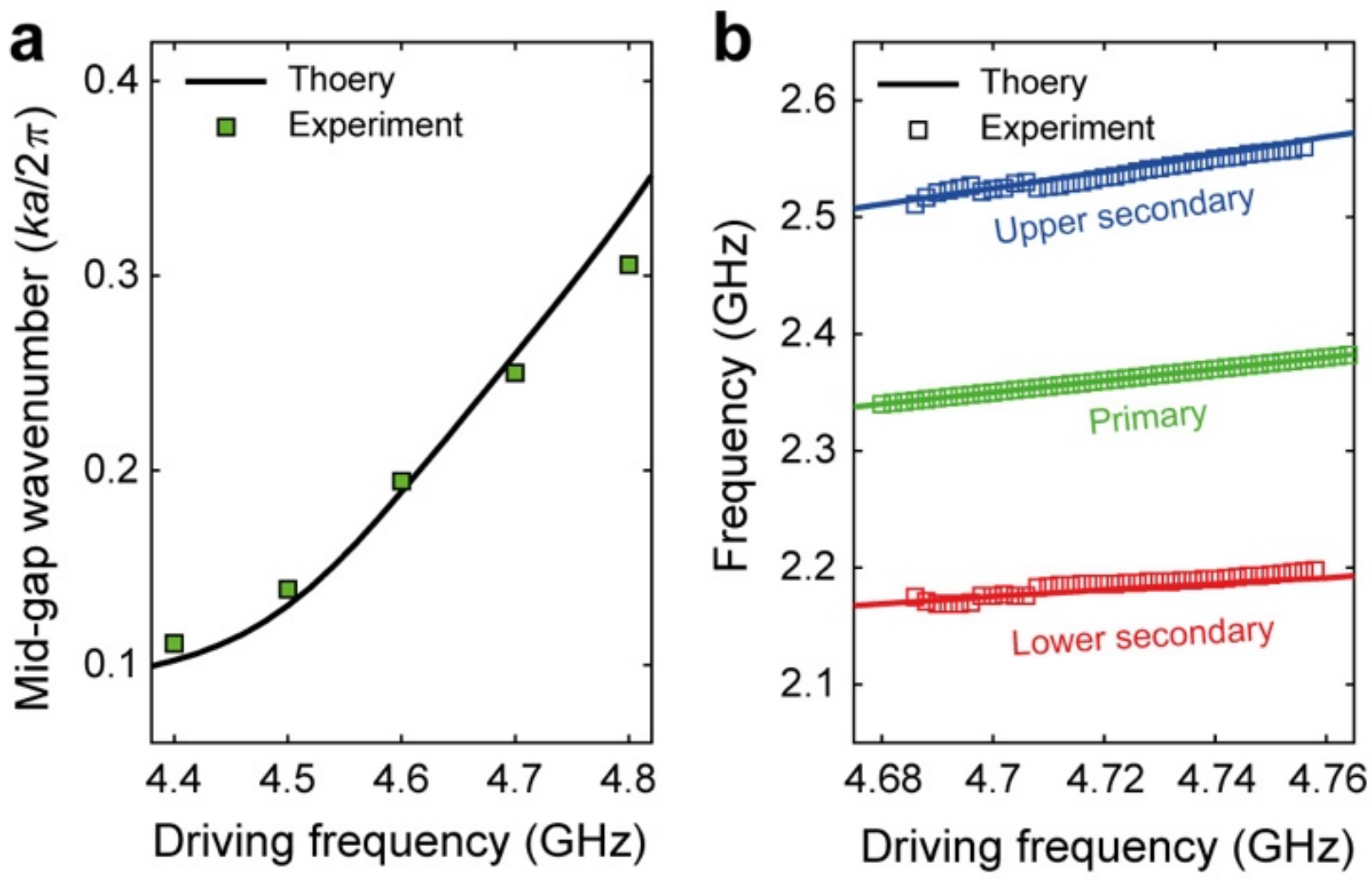

Fig. 3 Park et al.

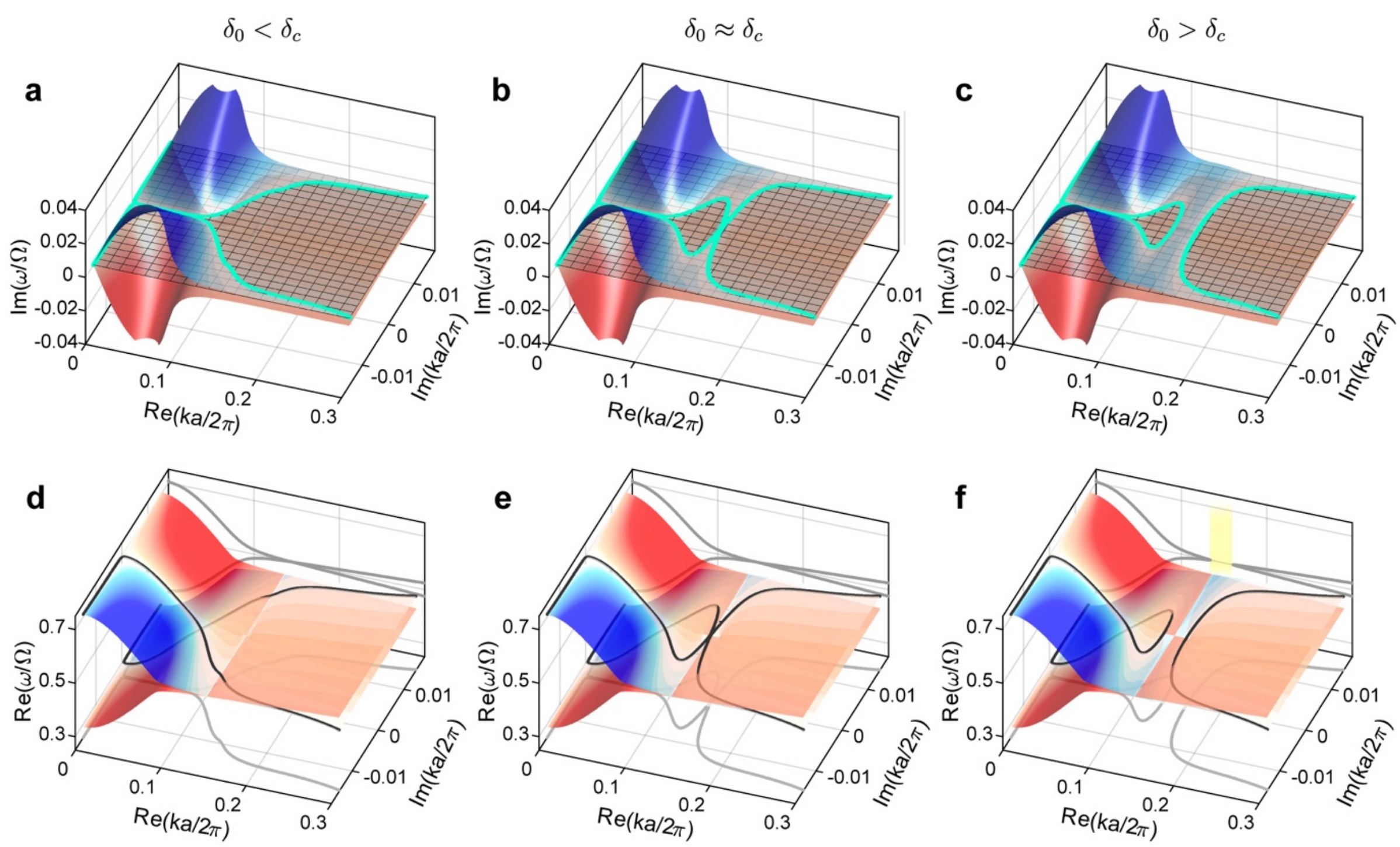

Fig. 4 Park et al.

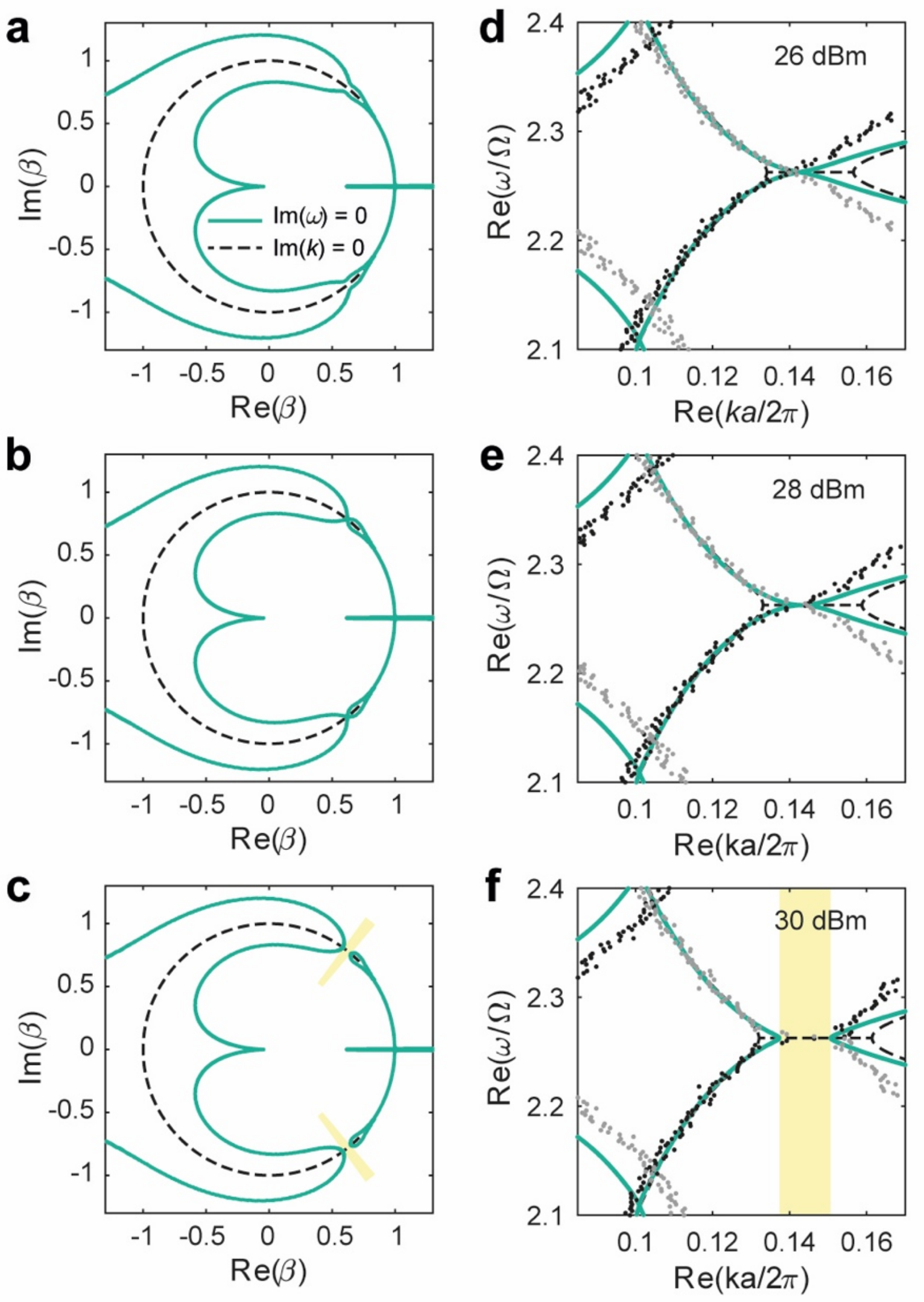

Fig. 5 Park et al.

# Supplementary Information for “Revealing non-Hermitian band structures of photonic Floquet media”

**Jagang Park[1,†], Hyukjoon Cho[1,†], Seojoo Lee[1†], Kyungmin Lee[1], Kanghee Lee[1], Hee Chul Park[2], Jung-Wan Ryu[2], Namkyoo Park[3], Sanggeun Jeon[4], and Bumki Min[1,*]**

[1]*Department of Mechanical Engineering, Korea Advanced Institute of Science and Technology (KAIST), Daejeon 34141, Republic of Korea*

[2]*Center for Theoretical Physics of Complex Systems, Institute for Basic Science (IBS) Daejeon 34126, Republic of Korea*

[3]*Department of Electrical and Computer Engineering, Seoul National University, Seoul 08826, Republic of Korea*

[4]*School of Electrical Engineering, Korea University, Seoul 02841, Republic of Korea*

In this supplementary information, we present details of theoretical analysis, sample fabrication, experimental setup and additional measurement for "Revealing non-Hermitian band structures of photonic Floquet media."

[*] Correspondence to: bmin@kaist.ac.kr

[†]These authors contributed equally to this work.

## 1. Transmission Line Modelling and Static Band Structure

We established a time-varying transmission line (TVTL) model with lumped circuit elements to analyze Floquet media, the properties of which are simultaneously dispersive and time varying. The model consists of a chain of driven LC resonators and a bus waveguide, as depicted in Fig. 1b in the main manuscript. The constituting resonators are assumed to be capacitively coupled to the waveguide and inductively coupled to their nearest neighbors. The Lagrangian can be given as a function of accumulated charge and electric current ($Q$ and $\dot{Q}$) at every nodal point and circuit element of the TVTL:

$$\mathcal{L} = \sum_n \left( \frac{1}{2} L_0 \left|\dot{Q}_n^{L_0}\right|^2 + \frac{1}{2} L_s \left|\dot{Q}_n^{L_s}\right|^2 + \frac{1}{2} L_r \left|\dot{Q}_n^{L_r}\right|^2 + \sum_i M_i \dot{Q}_n^{L_r} \dot{Q}_{n+i}^{L_r} - \frac{\left|Q_n^{C_0}\right|^2}{2C_0} - \frac{\left|Q_n^{C_r}\right|^2}{2C_r} - \frac{Q_n^{C_0} Q_n^{C_r}}{C_g} \right),$$

where $L_0$, $L_s$, $C_0$, $L_r$, $C_r$, $M_i$, and $C_g$ are the series inductance, shunt inductance and shunt capacitance of the bus waveguide, inductance and capacitance of the resonator, mutual inductance between $i$-th nearest-neighbor resonators and coupling capacitance between the resonator and the bus waveguide, respectively. The superscripts on $Q$s ($\dot{Q}$s) indicate the circuit elements conducting the electric current (storing electric charge).

The Kirchhoff's current law (KCL) and voltage law (KVL) impose following constraints:

(1) from the KCL, (C1) $\quad \dot{Q}_n^{L_0} = \dot{Q}_n^{L_s} + \dot{Q}_n^{C_s} + \dot{Q}_{n+1}^{L_0}$

(2) from the KVL, (C2) $\quad L_s \dot{Q}_{n-1}^{L_s} - L_s \dot{Q}_n^{L_s} - L_0 \dot{Q}_n^{L_0} = 0$

(3) from the KCL, (C3) $\quad \dot{Q}_n^{L_r} = \dot{Q}_n^{C_r}$

By taking discrete Fourier transform

$$Q_\kappa = \frac{1}{\sqrt{N}} \sum_n e^{-j(2\pi/N)\kappa n} Q_n,$$

we can rewrite the constraints in the momentum space as

(1) from C2: $\quad \dot{Q}_\kappa^{L_0} = -\frac{L_s}{L_0}\left(1 - e^{-j(2\pi/N)\kappa}\right)\dot{Q}_\kappa^{L_s} = -\frac{L_s}{L_0} p^* \dot{Q}_\kappa^{L_s}$

(2) from C1: $\quad \dot{Q}_\kappa^{C_s} = -\left(\frac{L_s}{L_0}|p|^2 + 1\right)\dot{Q}_\kappa^{L_s}$

where $p \equiv 1 - e^{j(2\pi/N)\kappa}$. Accordingly, the Lagrangian in the momentum space is found as

$$\mathcal{L} = \sum_{\kappa} \left[ \frac{L_s}{2} \left( \frac{L_s}{L_0} |p|^2 + 1 \right) \dot{Q}^{L_s}_{-\kappa} \dot{Q}^{L_s}_{\kappa} + \left( \frac{L_r}{2} + \sum_i M_i e^{j(2\pi/N)\kappa i} \right) \dot{Q}^{L_r}_{-\kappa} \dot{Q}^{L_r}_{\kappa} \right.$$

$$\left. - \left( \frac{L_s}{L_0} |p|^2 + 1 \right)^2 \frac{Q^{L_s}_{-\kappa} Q^{L_s}_{\kappa}}{2C_0} - \frac{Q^{L_r}_{-\kappa} Q^{L_r}_{\kappa}}{2C_r} + \left( \frac{L_s}{L_0} |p|^2 + 1 \right) \frac{Q^{L_s}_{-\kappa} Q^{L_s}_{\kappa}}{C_g} \right].$$

The canonical node flux (conjugate momentum of the charge, $\Phi_\kappa = \partial \mathcal{L} / \partial \dot{Q}_\kappa$) are found as

$$\Phi^{L_s}_{\kappa} = \frac{\partial \mathcal{L}}{\partial \dot{Q}^{L_s}_{\kappa}} = L_s \left( \frac{L_s}{L_0} |p|^2 + 1 \right) \dot{Q}^{L_s}_{-\kappa},$$

$$\Phi^{L_r}_{\kappa} = \frac{\partial \mathcal{L}}{\partial \dot{Q}^{L_r}_{\kappa}} = [L_r + 2 \textstyle\sum_i M_i \cos(2\pi/N)\kappa i] \dot{Q}^{L_r}_{-\kappa} \equiv L'_r \dot{Q}^{L_r}_{-\kappa},$$

and inversely,

$$\dot{Q}^{L_s}_{\kappa} = \frac{\Phi^{L_s}_{\kappa}}{L_s \left( \frac{L_s}{L_0} |p|^2 + 1 \right)}, \text{ and } \dot{Q}^{L_r}_{\kappa} = \frac{\Phi^{L_r}_{-\kappa}}{L'_r}.$$

The Legendre transformation of the Fourier transformed Lagrangian gives the Hamiltonian in the momentum space as a function of charge and node flux ($Q$ and $\Phi$), which are the canonical variables for electric circuits.

$$\mathcal{H} = \sum_{\kappa} \frac{\partial \mathcal{L}}{\partial \dot{Q}_\kappa} \dot{Q}_\kappa - \mathcal{L}$$

$$= \sum_{\kappa} \left( \frac{\Phi^{L_s}_{-\kappa} \Phi^{L_s}_{\kappa}}{2 L_s^2 C_0 \Omega_\kappa^2} + \frac{\Phi^{L_r}_{-\kappa} \Phi^{L_r}_{\kappa}}{2 L'_r} + \frac{1}{2} L_s^2 C_0 \Omega_\kappa^4 Q^{L_s}_{-\kappa} Q^{L_s}_{\kappa} + \frac{Q^{L_s}_{-\kappa} Q^{L_s}_{-\kappa}}{2C_r} - L_s C_0 \Omega_\kappa^2 \frac{Q^{L_s}_{-\kappa} Q^{L_r}_{\kappa} - Q^{L_r}_{-\kappa} Q^{L_s}_{\kappa}}{2C_g} \right)$$

Here, the frequency of the rectangular waveguide band is given to be

$$\Omega_\kappa \equiv \sqrt{\frac{1}{L_s C_0} \left( \frac{L_s}{L_0} |1 - e^{j(2\pi/N)\kappa}|^2 + 1 \right)} = \sqrt{\frac{4 \sin^2(2\pi/N)\kappa}{L_0 C_0} + \frac{1}{L_s C_0}}.$$

Defining the amplitudes of the waveguide mode ($a_\kappa$) and the coupled resonator mode ($b_\kappa$) as linear combinations of the canonical variables as below:

$$a_\kappa = \sqrt{\frac{L_s^2 C_0 \Omega_\kappa^3}{2}} Q^{L_s}_{\kappa} + j \sqrt{\frac{1}{2 L_s^2 C_0 \Omega_\kappa^3}} \Phi^{L_s}_{\kappa}, \text{ and } b_\kappa = \sqrt{\frac{1}{2 C_r \omega_\kappa}} Q^{L_r}_{\kappa} + j \sqrt{\frac{C_r \omega_\kappa}{2}} \Phi^{L_r}_{\kappa},$$

the Hamiltonian can be rewritten and divided into three parts as:

$$\mathcal{H} = \sum_\kappa \mathcal{H}_\kappa^a + \mathcal{H}_\kappa^b + \mathcal{H}_\kappa^{int} = \sum_\kappa \frac{\Omega_\kappa}{2}\left(a_\kappa^\dagger a_\kappa + a_{-\kappa}^\dagger a_{-\kappa}\right) + \frac{\omega_\kappa}{2}\left(b_\kappa^\dagger b_\kappa + b_{-\kappa}^\dagger b_{-\kappa}\right) - \frac{g_\kappa}{2}\left[\left(a_{-\kappa} + a_\kappa^\dagger\right)\left(b_\kappa + b_{-\kappa}^\dagger\right) + \left(b_{-\kappa} + b_\kappa^\dagger\right)\left(a_\kappa + a_{-\kappa}^\dagger\right)\right].$$

Here, $\omega_\kappa \equiv \sqrt{1/L_r'C_r}$ is the frequency of the coupled-resonator band, and $g_\kappa \equiv \sqrt{C_0 C_r \Omega_\kappa \omega_\kappa}/2C_g$ is the coupling coefficient between the waveguide and the resonators.

### (1) Rotating-wave approximation

Under the rotating-wave approximation (RWA), the Hamiltonian can be reduced to

$$\mathcal{H} = \sum_\kappa \Omega_\kappa a_\kappa^\dagger a_\kappa + \omega_\kappa b_\kappa^\dagger b_\kappa - g_\kappa\left(a_\kappa^\dagger b_\kappa + b_\kappa^\dagger a_\kappa\right),$$

which can be rewritten in a compact form on the basis of mode amplitudes $\mathbf{x} = (a_\kappa, b_\kappa)^T$ as

$$\mathcal{H} = \sum_\kappa \mathbf{x}^\dagger \widehat{\mathbf{H}}_\kappa \mathbf{x},$$

where

$$\widehat{\mathbf{H}}_\kappa = \begin{bmatrix} \Omega_\kappa & -g_\kappa \\ -g_\kappa & \omega_\kappa \end{bmatrix}.$$

The static (time-invariant) band structure of non-driven photonic media can be calculated from the eigenvalues of the Hamiltonian matrix as

$$\omega_1 = \frac{1}{2}\left[(\Omega_\kappa + \omega_\kappa) - \sqrt{(\Omega_\kappa - \omega_\kappa)^2 + 4g_\kappa^2}\right],$$

$$\omega_2 = \frac{1}{2}\left[(\Omega_\kappa + \omega_\kappa) + \sqrt{(\Omega_\kappa - \omega_\kappa)^2 + 4g_\kappa^2}\right],$$

which represents two bands originating from avoided crossing of the waveguide band and the coupled resonator band. These approximated static band structures are drawn in Fig. 1c in the main manuscript.

### (2) Beyond the rotating-wave approximation

Without the rotating-wave approximation, the Hamiltonian can be written in a compact form on the basis of four mode amplitudes $\mathbf{x} = \left(a_\kappa, b_\kappa, a_{-\kappa}^\dagger, b_{-\kappa}^\dagger\right)^T$ as

$$\hat{\mathbf{H}}_\kappa = \frac{1}{2}\begin{bmatrix} \Omega_\kappa & -g_\kappa & 0 & -g_\kappa \\ -g_\kappa & \omega_\kappa & -g_\kappa & 0 \\ 0 & -g_\kappa & \Omega_\kappa & -g_\kappa \\ -g_\kappa & 0 & -g_\kappa & \omega_\kappa \end{bmatrix}.$$

The Hamilton's equation can be derived as

$$j\frac{\partial \mathbf{x}}{\partial t} = j\mathbf{MJM}^\dagger \frac{\partial \mathcal{H}}{\partial \mathbf{x}} \equiv H_\kappa \mathbf{x} = \begin{bmatrix} \Omega_\kappa & -g_\kappa & 0 & -g_\kappa \\ -g_\kappa & \omega_\kappa & -g_\kappa & 0 \\ 0 & g_\kappa & -\Omega_\kappa & g_\kappa \\ g_\kappa & 0 & g_\kappa & -\omega_\kappa \end{bmatrix}\mathbf{x},$$

where $\mathbf{M}$ is a linear transformation matrix such that $\mathbf{M}\left[Q_\kappa^{L_s}, Q_\kappa^{L_r}, \Phi_\kappa^{L_s}, \Phi_\kappa^{L_r}\right]^T = \mathbf{x}$, and $\mathbf{J} = -j\sigma_y \otimes \mathbf{I}_2$. The band structure can be obtained by calculating the eigenvalues of $H_\kappa$ for each $\kappa$,

$$\omega_1^2 = \frac{1}{2}\left[(\Omega_\kappa^2 + \omega_\kappa^2) - \sqrt{(\Omega_\kappa^2 - \omega_\kappa^2)^2 + 16\omega_\kappa\Omega_\kappa g_\kappa^2}\right],$$

$$\omega_2^2 = \frac{1}{2}\left[(\Omega_\kappa^2 + \omega_\kappa^2) + \sqrt{(\Omega_\kappa^2 - \omega_\kappa^2)^2 + 16\omega_\kappa\Omega_\kappa g_\kappa^2}\right].$$

Note that these eigenvalues are reduced to those of rotating-wave approximated ones for sufficiently small coupling coefficient $g_\kappa$.

## 2. Inclusion of Dissipative Loss in Resonators

The presence of intrinsic loss makes the eigenfrequencies of a system basically complex-valued, and it is responsible for the visibility of the momentum gap in the phase retardation measurements as shown in Figs. 4 and 5 in the main manuscript. Considering our experimental platform consisting of varactor-loaded *LC*-resonators, the dominant intrinsic loss channel would be the Ohmic loss by the current leakage through the varactor diodes. We modeled the Ohmic loss by adding a series resistor $R$ to each resonator, then defined the Rayleigh dissipation function as a function of the electric current in the resonators:

$$\mathcal{F}(\dot{Q}) = \sum_{\kappa} \frac{1}{2} R \left| \dot{Q}_{\kappa}^{L_r} \right|^2 = \sum_{\kappa} \frac{1}{2} \frac{R}{L_r'^2} \left| \Phi_{\kappa}^{L_r} \right|^2 = -\sum_{\kappa} \frac{R\omega_{\kappa}}{4L_r'} \left( b_{-\kappa} - b_{\kappa}^{\dagger} \right) \left( b_{\kappa} - b_{-\kappa}^{\dagger} \right),$$

or in the compact form $\mathcal{F} = \sum_{\kappa} \mathbf{x}^{\dagger} \hat{\mathbf{F}}_{\kappa} \mathbf{x}$,

$$\hat{F}_{\kappa} = \frac{R\omega_{\kappa}}{4L_r'} \begin{bmatrix} 0 & 0 & 0 & 0 \\ 0 & 1 & 0 & -1 \\ 0 & 0 & 0 & 0 \\ 0 & -1 & 0 & 1 \end{bmatrix}.$$

The Rayleigh dissipation function gives the non-conservative damping force term $\mathbf{f}_{nc} = -\partial\mathcal{F}/\partial\dot{\mathbf{Q}}$, which is added to the Hamilton's equation. With some notational modifications, the non-conservative force term becomes

$$\begin{bmatrix} \mathbf{0} \\ \mathbf{f}_{nc} \end{bmatrix} = \begin{bmatrix} 0 & 0 \\ 0 & 1 \end{bmatrix} \begin{bmatrix} -\partial\mathcal{F}/\partial\mathbf{Q} \\ -\partial\mathcal{F}/\partial\dot{\mathbf{Q}} \end{bmatrix} = \begin{bmatrix} 0 & 0 & 0 & 0 \\ 0 & 0 & 0 & 0 \\ 0 & 0 & L_r' & 0 \\ 0 & 0 & 0 & L_r' \end{bmatrix} \begin{bmatrix} -\partial\mathcal{F}/\partial\mathbf{Q} \\ -\partial\mathcal{F}/\partial\mathbf{\Phi} \end{bmatrix} \equiv -\mathbf{L}\frac{\partial\mathcal{F}}{\partial\boldsymbol{\eta}},$$

which leads to the dissipative Hamiltonian matrix

$$H_{\kappa} = \begin{bmatrix} \Omega_{\kappa} & g_{\kappa} & 0 & -g_{\kappa} \\ g_{\kappa} & \omega_{\kappa} & -g_{\kappa} & 0 \\ 0 & g_{\kappa} & -\Omega_{\kappa} & -g_{\kappa} \\ g_{\kappa} & 0 & -g_{\kappa} & -\omega_{\kappa} \end{bmatrix} - j\gamma \begin{bmatrix} 0 & 0 & 0 & 0 \\ 0 & 1 & 0 & -1 \\ 0 & 0 & 0 & 0 \\ 0 & -1 & 0 & 1 \end{bmatrix},$$

where the loss rate is defined as $\gamma \equiv R/4L_r'$. Under the rotating-wave approximation, complex-valued four eigenvalues are obtained as follows:

$$\omega_{1+} = \frac{1}{2}\left[ (\Omega_{\kappa} + \omega_{\kappa} - j\gamma) - \sqrt{(\Omega_{\kappa} - \omega_{\kappa} + j\gamma)^2 + g_{\kappa}^2} \right],$$

$$\omega_{2+} = \frac{1}{2}\left[(\Omega_\kappa + \omega_\kappa - j\gamma) + \sqrt{(\Omega_\kappa - \omega_\kappa + j\gamma)^2 + g_\kappa^2}\right],$$

$$\omega_{1-} = \frac{1}{2}\left[(-\Omega_\kappa - \omega_\kappa - j\gamma) + \sqrt{(\Omega_\kappa - \omega_\kappa - j\gamma)^2 + g_\kappa^2}\right] = -\omega_{1+}^*,$$

$$\omega_{2-} = \frac{1}{2}\left[(-\Omega_\kappa - \omega_\kappa - j\gamma) - \sqrt{(\Omega_\kappa - \omega_\kappa - j\gamma)^2 + g_\kappa^2}\right] = -\omega_{2+}^*.$$

The loss-considered complex-valued band structures were used to interpret the phase retardation measurements and the visibility of the momentum band gap as shown in Figs. 4 and 5 in the main manuscript.

## 3. Theoretical Calculation of Floquet Band Structure

### (1) Derivation of Floquet Hamiltonian

For the calculation of Bloch-Floquet band structures, we first assumed that the resonator capacitor is driven time-periodically, such that the capacitance is given to be

$$C_r(t) = \frac{C_c}{1+\delta(t)},$$

with a temporal periodicity of $T = 2\pi/\Omega$; this assumption makes the Hamiltonian time-periodic, i.e., $\mathcal{H}(t+2\pi/\Omega) = \mathcal{H}(t)$. Especially, $\mathcal{H}_\kappa^b$ terms in the Hamiltonian that are directly dependent on the resonator capacitance becomes

$$\mathcal{H}_\kappa^b(t) = \frac{\omega_\kappa}{2}\left(b_\kappa^\dagger b_\kappa + b_{-\kappa}^\dagger b_{-\kappa}\right) + \frac{\omega_\kappa \delta(t)}{4}\left(b_{-\kappa} + b_\kappa^\dagger\right)\left(b_\kappa + b_{-\kappa}^\dagger\right).$$

In the compact form,

$$\widehat{\mathbf{H}}_\kappa(t) = \widehat{H}_\kappa^0 + \widehat{H}_\kappa^{TV}(t) = \frac{1}{2}\begin{bmatrix} \Omega_\kappa & -g_\kappa & 0 & -g_\kappa \\ -g_\kappa & \omega_\kappa & -g_\kappa & 0 \\ 0 & -g_\kappa & \Omega_\kappa & -g_\kappa \\ -g_\kappa & 0 & -g_\kappa & \omega_\kappa \end{bmatrix} + \frac{\omega_\kappa \delta(t)}{4}\begin{bmatrix} 0 & 0 & 0 & 0 \\ 0 & 1 & 0 & 1 \\ 0 & 0 & 0 & 0 \\ 0 & 1 & 0 & 1 \end{bmatrix}.$$

The final Hamilton's equation with the time-periodic, dissipative Hamiltonian matrix with is found as

$$j\frac{\partial \psi_\kappa(t)}{\partial t} = H_\kappa(t)\psi_\kappa(t),$$

where

$$H_\kappa(t) = \begin{bmatrix} \Omega_\kappa & -g_\kappa & 0 & -g_\kappa \\ -g_\kappa & \omega_\kappa - j\gamma & -g_\kappa & 0 \\ 0 & g_\kappa & -\Omega_\kappa & g_\kappa \\ g_\kappa & 0 & g_\kappa & -\omega_\kappa - j\gamma \end{bmatrix} + \frac{\omega_\kappa \delta(t)}{2}\begin{bmatrix} 0 & 0 & 0 & 0 \\ 0 & 1 & 0 & 1 \\ 0 & 0 & 0 & 0 \\ 0 & -1 & 0 & -1 \end{bmatrix}.$$

To construct effective Floquet Hamiltonian matrix, we Fourier expanded the time-periodic Hamiltonian matrix in terms of $e^{-j\Omega t}$ as

$$H_\kappa(t) = \textstyle\sum_q e^{-jq\Omega t}\widetilde{H}_\kappa^q,$$

which allows us to choose the Floquet solution as

$$\psi_\kappa(t) = e^{-j\omega t}\sum_q e^{-jq\Omega t}\tilde{\psi}_\kappa^q.$$

Putting these series into the Hamilton's equation,

$$j\sum_q -j(\omega+q\Omega)e^{-j(\omega+q\Omega)t}\tilde{\psi}_\kappa^q = \left(\sum_q e^{-jq\Omega t}\tilde{H}_\kappa^q\right)\left(\sum_q e^{-j(\omega+q\Omega)t}\tilde{\psi}_\kappa^q\right)$$

$$= \sum_q\sum_r e^{-j(q-r)\Omega t}\tilde{H}_\kappa^{q-r}e^{-j(\omega+r\Omega)t}\tilde{\psi}_\kappa^r = \sum_q e^{-j(\omega+q\Omega)t}\sum_r \tilde{H}_\kappa^{q-r}\tilde{\psi}_\kappa^r.$$

Based on the orthogonality of the Harmonic terms, the Hamilton's equation can be rearranged into a set of time-independent equations,

$$(\omega+q\Omega)\tilde{\psi}_\kappa^q = \sum_r \tilde{H}_\kappa^{q-r}\tilde{\psi}_\kappa^r.$$

The set of equations can be recast into an eigenvalue problem with following time-independent effective Floquet Hamiltonian matrix;

$$H_\kappa^F = \begin{bmatrix} \ddots & & & & \\ & \tilde{H}_\kappa^0+\Omega\mathbf{I}_4 & \tilde{H}_\kappa^{-1} & \tilde{H}_\kappa^{-2} & \\ & \tilde{H}_\kappa^{+1} & \tilde{H}_\kappa^0 & \tilde{H}_\kappa^{-1} & \\ & \tilde{H}_\kappa^{+2} & \tilde{H}_\kappa^{+1} & \tilde{H}_\kappa^0-\Omega\mathbf{I}_4 & \\ & & & & \ddots \end{bmatrix}.$$

The size of the Floquet Hamiltonian matrix depends on the harmonic order considered in the analysis and determines the number of shifted replicas of static bands appearing in the band structure. In the case where $\delta(t)$ is sinusoidal, i.e., $\delta(t) = \delta_0\cos(\Omega t+\phi_0)$, we can truncate the Floquet Hamiltonian matrix up to the first harmonic order without significant loss of accuracy. The calculated Bloch-Floquet band structures are shown in Fig. 1d and overlaid on the experimentally measured band structure in Fig. 2 in the main manuscript. In the following subsection, we will diagonalize the zeroth-order Hamiltonian, so that we could get a clearer picture of the time-modulation-induced interaction between the bands.

### (2) Reduced Floquet Hamiltonian matrix with rotating-wave approximation

Opening of a gap along the momentum axis is attributed to the non-reciprocal coupling between two dominant bands participating in the interaction, one of which is the undressed (static) band and the other is the dressed band. Therefore, the driving-induced interaction between the two bands can be intuitively understood by constructing a 2×2 effective Hamiltonian matrix, so-called the reduced Floquet Hamiltonian. Assuming that the capacitance driving is simply sinusoidal as $\delta(t) = \delta_0 \cos(\Omega t + \phi_0)$, the Fourier series for the Hamiltonian matrix can be truncated up to their first harmonic terms,

$$\widetilde{H}_\kappa^{\pm 1} = \frac{\omega_\kappa \delta_0}{4} e^{\mp j\phi_0} \begin{bmatrix} 0 & 0 & 0 & 0 \\ 0 & 1 & 0 & 1 \\ 0 & 0 & 0 & 0 \\ 0 & -1 & 0 & -1 \end{bmatrix}.$$

To ease the algebraic complexity in the diagonalization, here we assume the waveguide-resonator coupling coefficient $g_\kappa$ is relatively small so that the RWA is valid. Then we have

$$\widetilde{H}_\kappa^0 \approx \begin{bmatrix} \Omega_\kappa & -g_\kappa & 0 & 0 \\ -g_\kappa & \omega_\kappa - j\gamma & 0 & 0 \\ 0 & 0 & -\Omega_\kappa & g_\kappa \\ 0 & 0 & g_\kappa & -\omega_\kappa - j\gamma \end{bmatrix},$$

which can be diagonalized as $P^{-1}\widetilde{H}_\kappa^0 P = \mathrm{diag}(\omega_{1+}, \omega_{2+}, \omega_{1-}, \omega_{2-})$ by a block diagonal matrix *P*:

$$P = \begin{bmatrix} 1 & 1 & 0 & 0 \\ \dfrac{\Omega_\kappa - \omega_{1+}}{g_\kappa} & \dfrac{\Omega_\kappa - \omega_{2+}}{g_\kappa} & 0 & 0 \\ 0 & 0 & 1 & 1 \\ 0 & 0 & \dfrac{\Omega_\kappa + \omega_{1-}}{g_\kappa} & \dfrac{\Omega_\kappa + \omega_{2-}}{g_\kappa} \end{bmatrix} \equiv \begin{bmatrix} 1 & 1 & 0 & 0 \\ \Delta_1 & \Delta_2 & 0 & 0 \\ 0 & 0 & 1 & 1 \\ 0 & 0 & \Delta_1^* & \Delta_2^* \end{bmatrix}.$$

By the diagonalization of the zeroth-order Hamiltonian matrix, the first-order harmonic components of the Hamiltonian matrix become

$$P^{-1}\widetilde{H}_\kappa^{\pm 1} P = \frac{\omega_\kappa \delta_0}{4} e^{\mp j\phi_0} \begin{bmatrix} \dfrac{1}{\Delta_2 - \Delta_1} \begin{bmatrix} -\Delta_1 & -\Delta_2 & -\Delta_1^* & -\Delta_2^* \\ \Delta_1 & \Delta_2 & \Delta_1^* & \Delta_2^* \end{bmatrix} \\ \dfrac{1}{\Delta_2^* - \Delta_1^*} \begin{bmatrix} \Delta_1 & \Delta_2 & \Delta_1^* & \Delta_2^* \\ -\Delta_1 & -\Delta_2 & -\Delta_1^* & -\Delta_2^* \end{bmatrix} \end{bmatrix}.$$

In this work, we mainly focus on the interaction between the positive fundamental band ($\omega_{1+}$) and the first harmonic of the negative fundamental band ($-\omega_{1-}+\Omega$). Finally, the 2×2 reduced Floquet Hamiltonian matrix is obtained by choosing only four elements from the entire Floquet Hamiltonian matrix that are associated with the driving-induced interaction between the involved bands.

$$H_{red}^{F}=\begin{bmatrix} \omega_{1-}+\Omega & \frac{\Delta_1}{\Delta_2^*-\Delta_1^*}\frac{\omega_\kappa\delta_0}{4}e^{j\phi_0} \\ \frac{-\Delta_1^*}{\Delta_2-\Delta_1}\frac{\omega_\kappa\delta_0}{4}e^{-j\phi} & \omega_{1+} \end{bmatrix},$$

and the nonreciprocal coupling coefficient for the dressed $\mu$-th band. and the undressed $\nu$-th band is found to be

$$\kappa_{\mu\nu}^{F}=\frac{\Delta_\nu}{\Delta_2^*-\Delta_1^*}\frac{\omega_\kappa\delta_0}{4}e^{j\phi_0},$$

where $\Delta_{\mu,\nu}\equiv\Omega_\kappa-\omega_{\mu+,\nu+}$, and $\mu,\nu\in\{1,2\}$. For the lossless case, the fundamental momentum gap created by the non-reciprocal interaction between positive and negative fundamental bands can be described by following reduced Floquet Hamiltonian;

$$H_{red}^{F}=\begin{bmatrix} \omega_{1-}+\Omega & -\frac{1}{2}\left(\frac{\Omega_\kappa-\omega_\kappa}{\sqrt{(\Omega_\kappa-\omega_\kappa)^2+4g_\kappa^2}}+1\right)\frac{\omega_\kappa\delta_0}{4}e^{j\phi_0} \\ \frac{1}{2}\left(\frac{\Omega_\kappa-\omega_\kappa}{\sqrt{(\Omega_\kappa-\omega_\kappa)^2+4g_\kappa^2}}+1\right)\frac{\omega_\kappa\delta_0}{4}e^{-j\phi_0} & \omega_{1+} \end{bmatrix}.$$

This matrix can be rewritten also as

$$H_{red}^{F}=\begin{bmatrix} \frac{\Omega}{2} & 0 \\ 0 & \frac{\Omega}{2} \end{bmatrix}+\begin{bmatrix} -\omega_{1+}^*+\frac{\Omega}{2} & \kappa_{11}^{F} \\ -\kappa_{11}^{F*} & \omega_{1+}-\Omega/2 \end{bmatrix}=\frac{\Omega}{2}\mathbf{I}_2+H_{APT}^{F}.$$

It is straightforward to verify that the latter matrix is anti-$\mathcal{PT}$ symmetric, i.e., $\{H_{APT}^{F},\mathcal{PT}\}=0$, for the conventional choice of the parity operator $\mathcal{P}=\sigma_x$. It is worth noting that the dissipative loss we considered in the previous section does not break the anti-$\mathcal{PT}$ symmetry, while it does affect the observability of the momentum gap in the phase retardation measurements.

**Supplementary Figures**

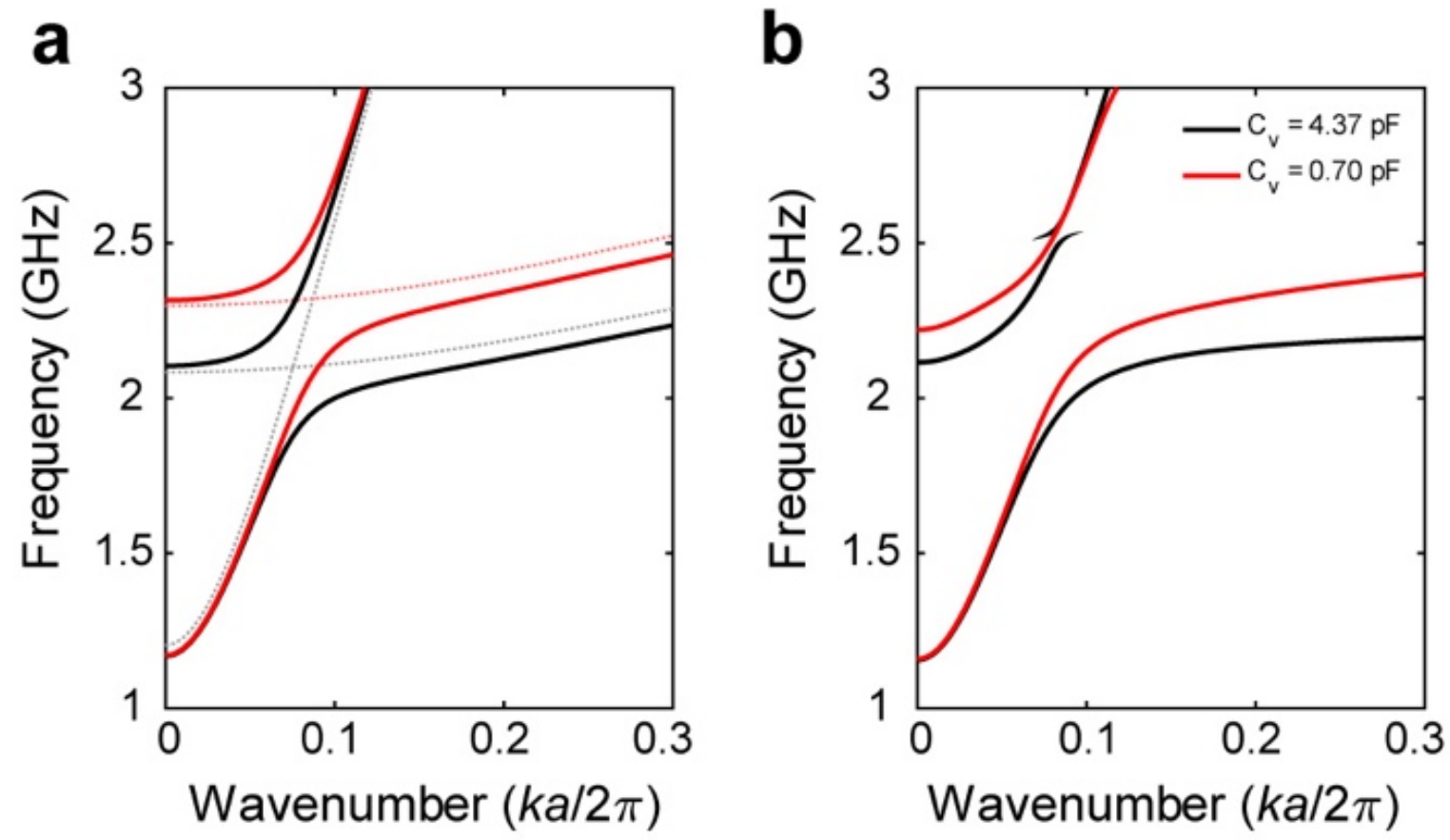

**Supplementary Figure S1. Static band structures of photonic Floquet media.** (a) Static band structure of photonic Floquet media calculated from the TVTL model for two different values of varactor capacitance, i.e., $\mathrm{C_v} = 4.37$ pF (drawn with black lines) and $\mathrm{C_v} = 0.70$ pF (drawn with red lines). The dotted lines denote the fundamental resonator and waveguide modes at each varactor capacitance value. (b) Static band structure of photonic Floquet media calculated from a finite element method (FEM) analysis for two different values of varactor capacitance. The band repulsion observed at ~2.5 GHz for $\mathrm{C_v} = 4.37$ pF is attributed to the interaction with a higher-order resonator mode.

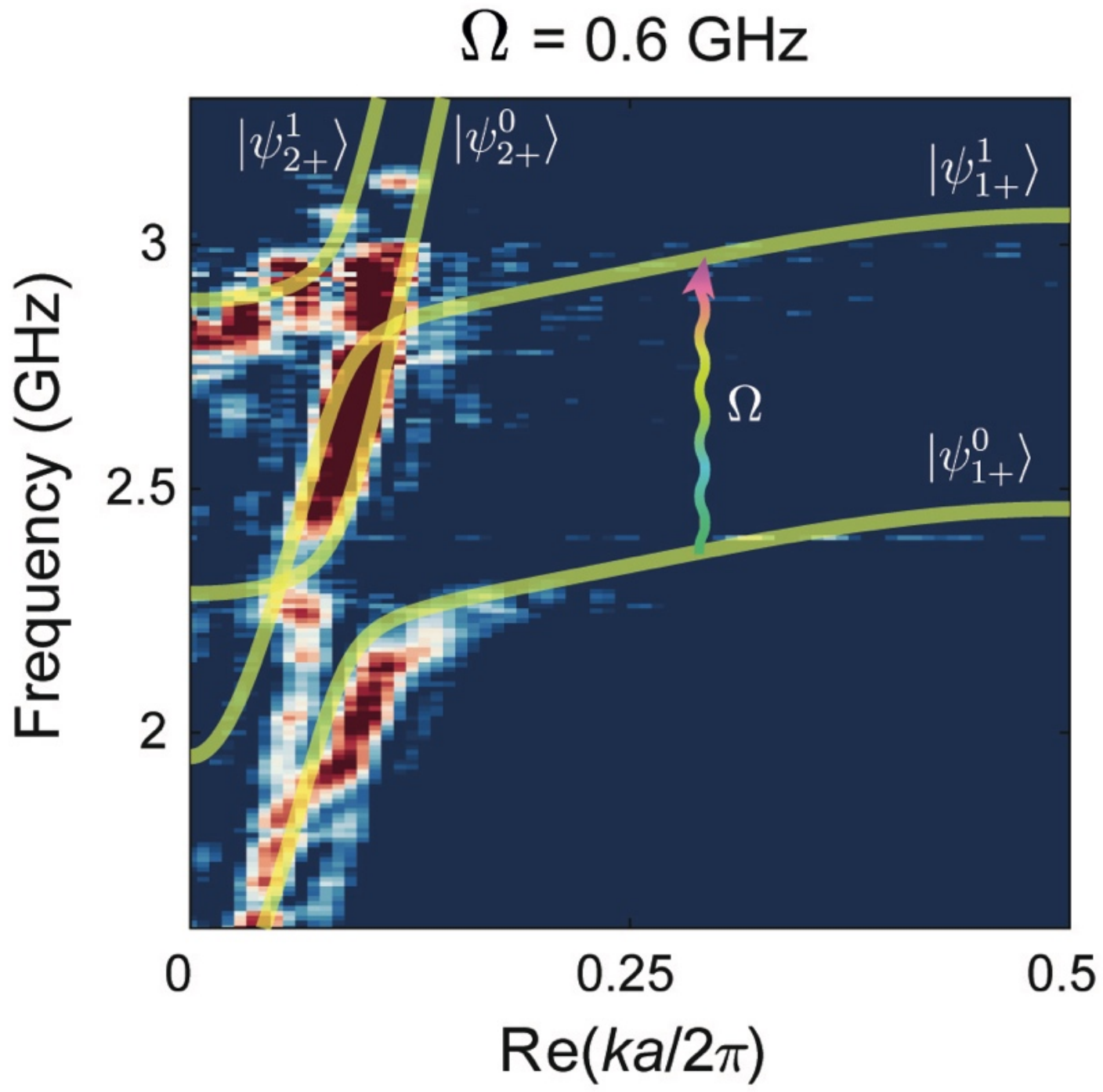

**Supplementary Figure S2. Observation of driving-induced dressing of positive frequency states.** Reconstructed Bloch-Floquet band structure of the photonic Floquet medium driven at 0.6 GHz. The theoretically calculated band structure is overlaid on the band structure (drawn with violet lines). The emergence of driving-dressed states $|\psi^1_{1+}\rangle$ and $|\psi^1_{2+}\rangle$ from the positive frequency lower bands can be observed.

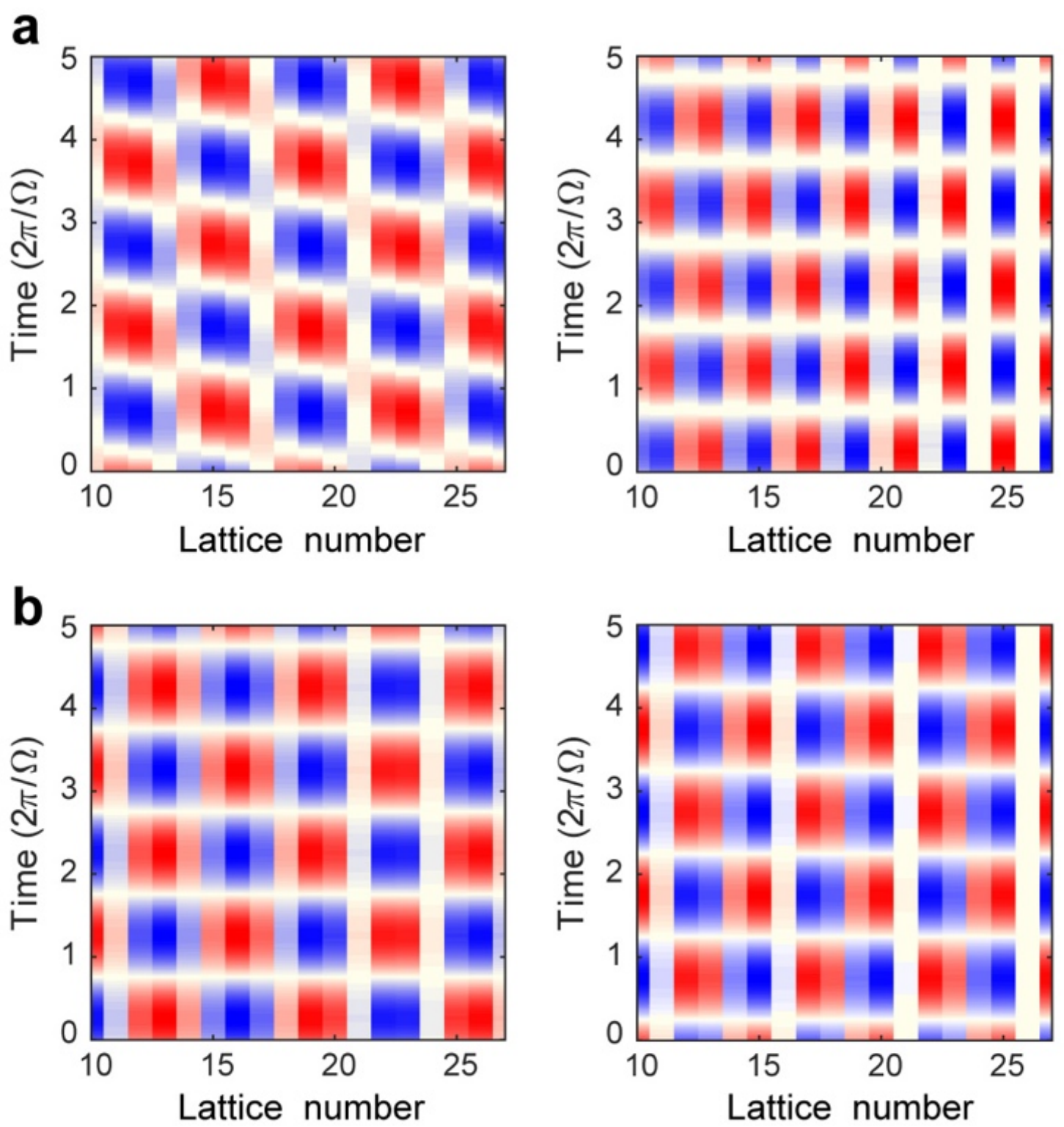

**Supplementary Figure S3. Mode field patterns at the edges of the primary momentum gap.** (a) Experimentally measured and (b) theoretically calculated spatiotemporal mode field patterns at the left (left panels) and right (right panels) edges of the momentum gap for the driving frequency of 4.65 GHz. Due to the limited spatial resolution, mode field patterns are acquired near the edges in the measurement, which leads to the discrepancy between the theoretically calculated and experimentally measured spatial field patterns. Nevertheless, a temporal phase shift of $\sim\pi/2$ between the mode field patterns at the left and right edges is clearly observed.

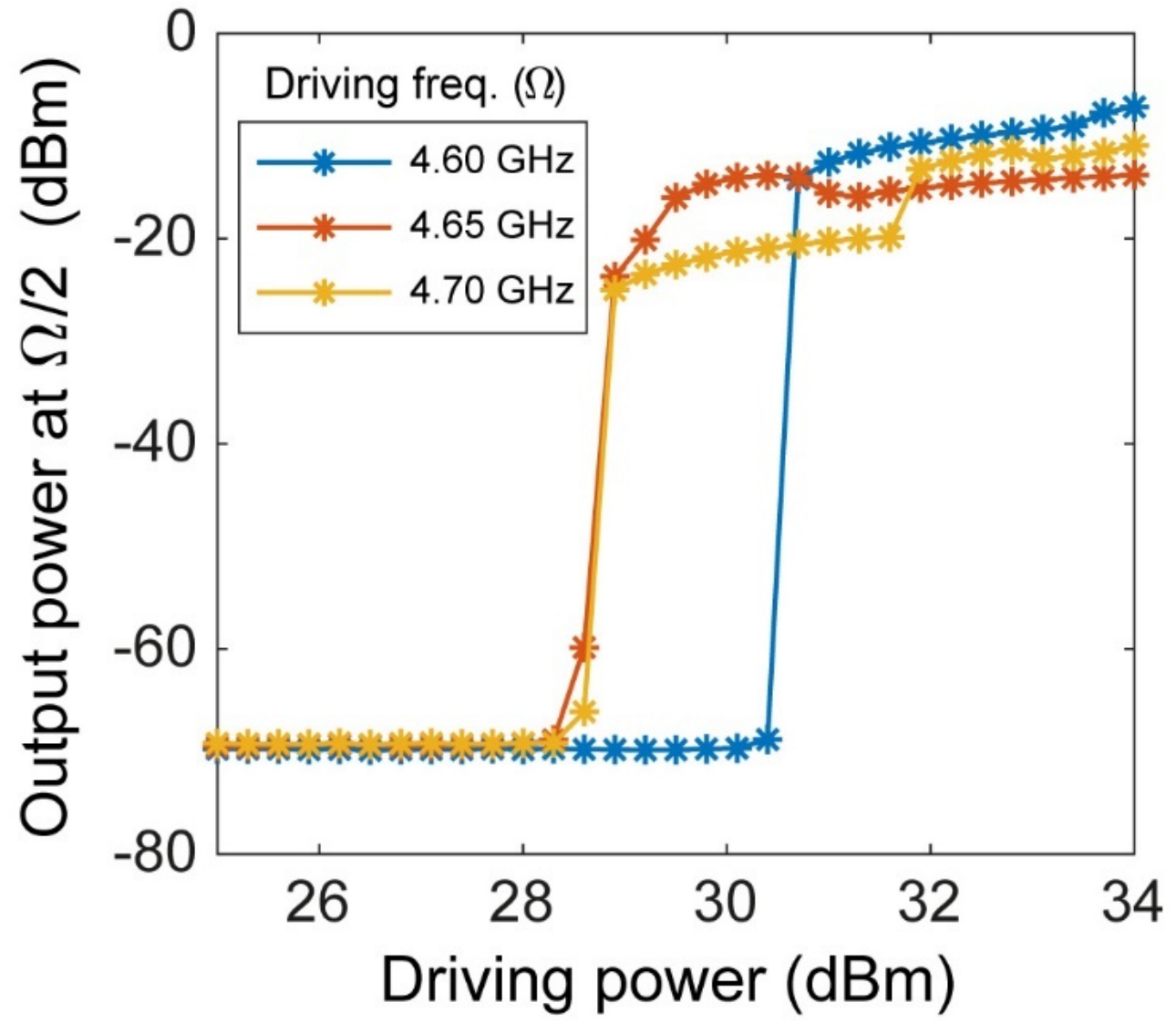

**Supplementary Figure S4. Characterisation of noise-initiated parametric oscillations.** Noise-initiated parametric oscillation power radiated from the photonic Floquet medium ($N$ = 24) plotted as a function of the driving power. Here, the driving frequencies were set to $\Omega/2\pi =$ 4.6 GHz, 4.65 GHz, and 4.7 GHz, and the driving power was monitored before being split to each of the unit cells. The oscillation threshold decreases as the driving frequency approaches twice the resonance frequency of the constituting resonator at the operating point.

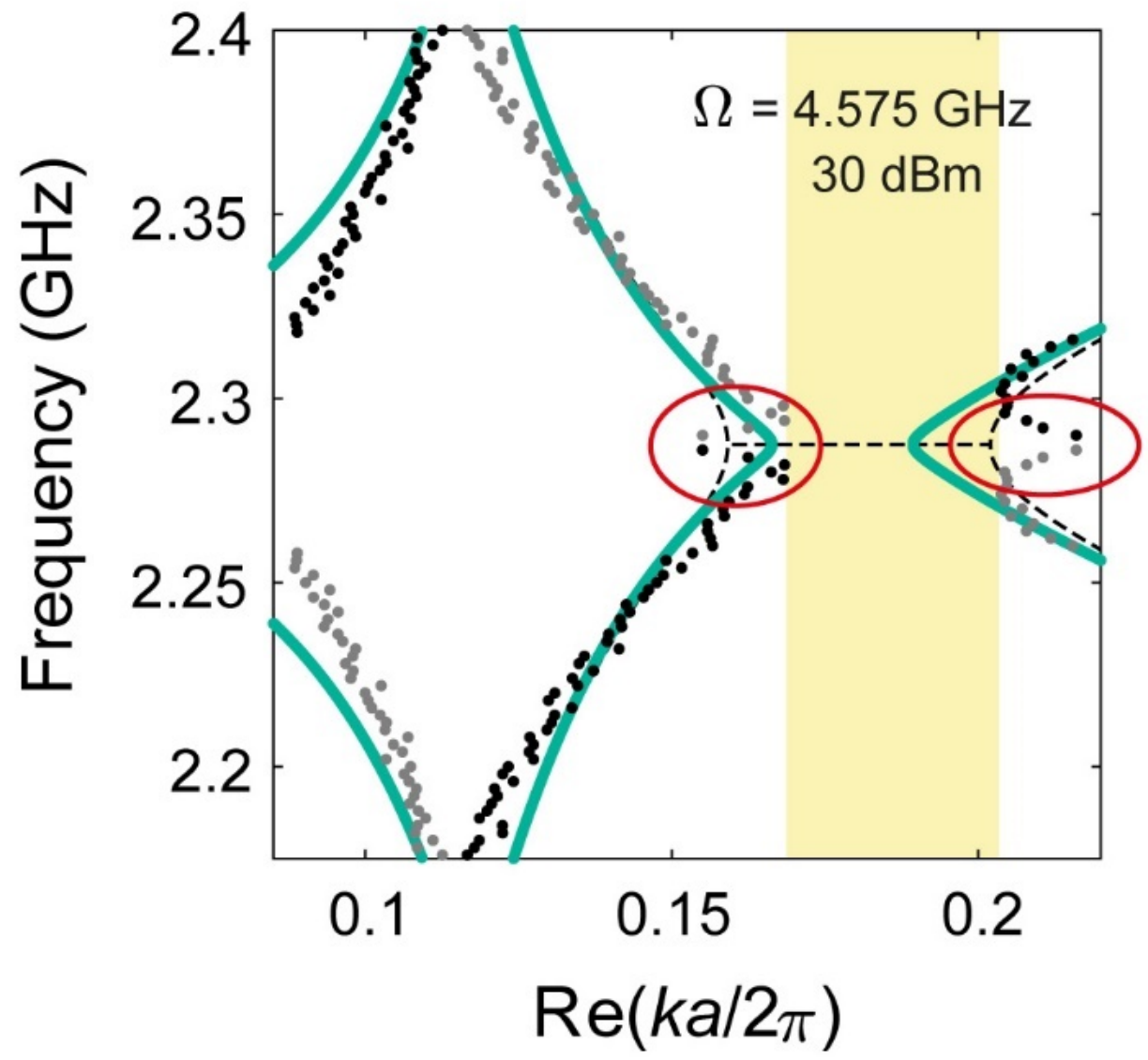

**Supplementary Figure 5. Invalidity of the single-pass approximation for the case of strong driving.** Reconstructed non-Bloch band structure for the case in which the single-pass approximation is not valid. In this driving regime, the value of phase retardation becomes exaggerated owing to the presence of multi-pass parametrically amplified waves in the photonic Floquet medium.

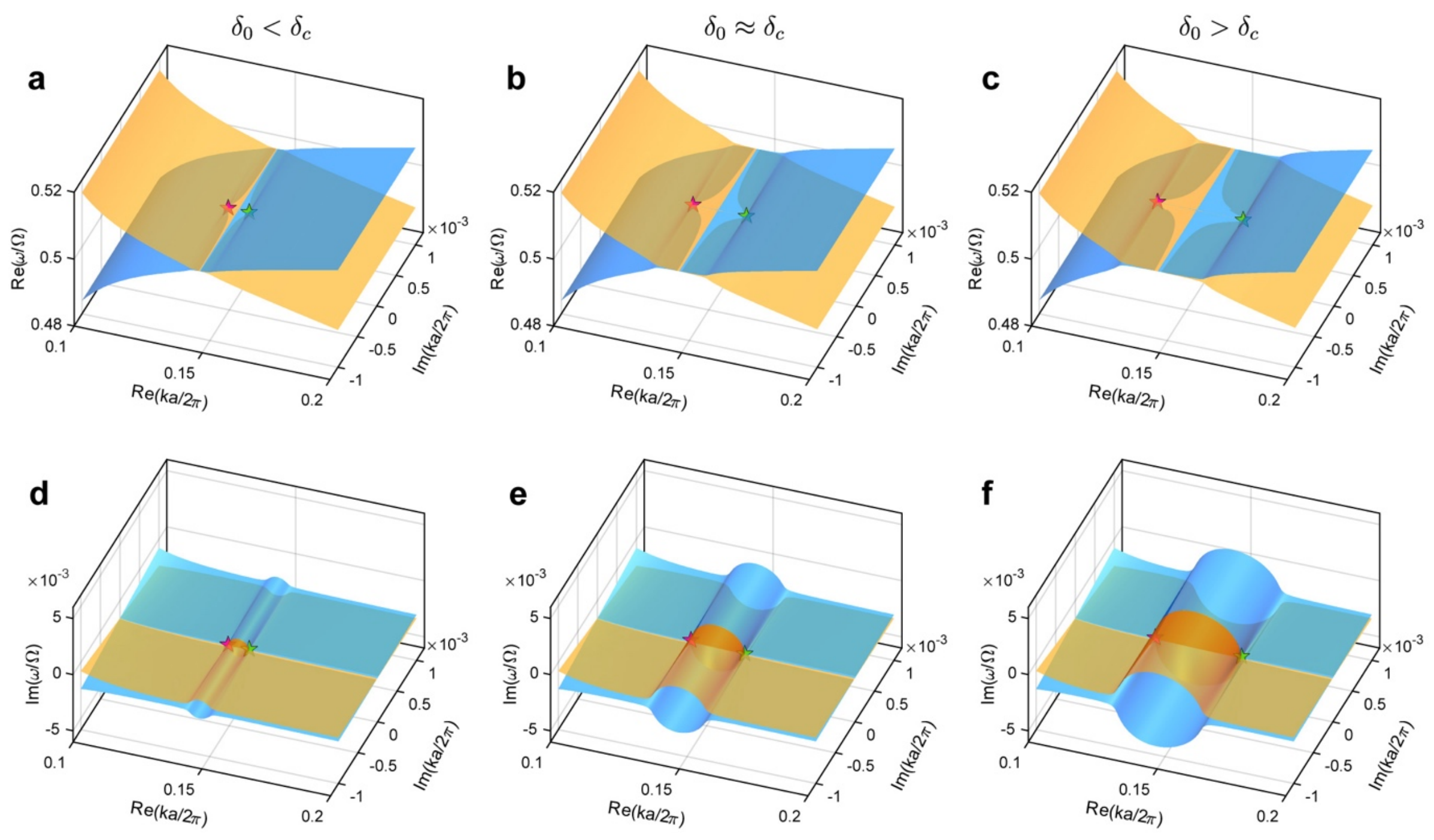

**Supplementary Figure S6. Expanded view of the complex eigenfrequency surfaces and their morphological change with a variation in the driving strength.** (a-c) Closer look at the real and (d-f) imaginary eigenfrequency surfaces near the primary momentum gap. The driving-induced morphological change of the complex eigenfrequency surfaces can be clearly observed in the plots in the region of complex momenta with real values approximately falling in the primary momentum gap. In each of the plots, two exceptional points are marked by red and green stars.

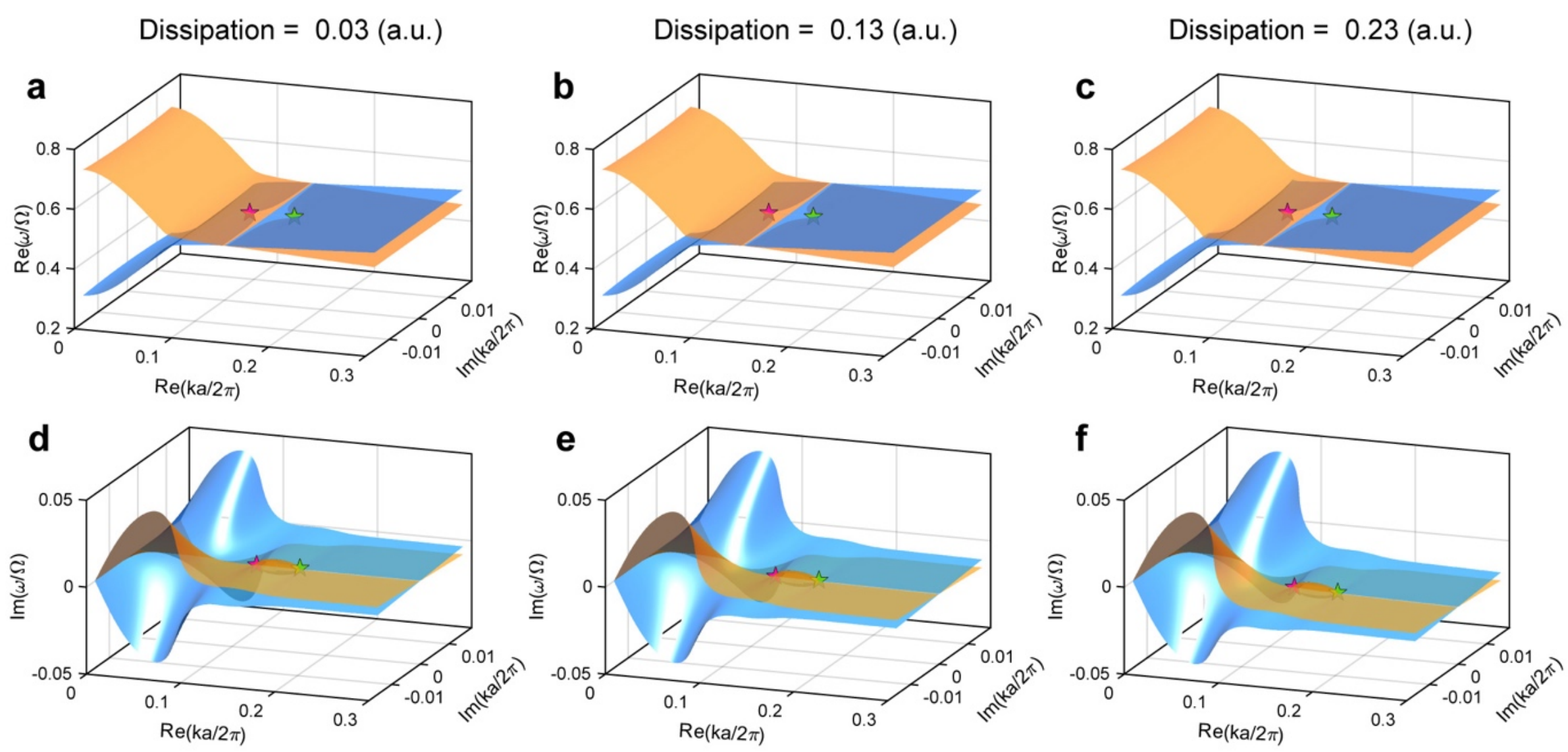

**Supplementary Figure S7. Morphological change of the complex eigenfrequency surfaces with a variation in the resonator dissipation rate.** Plots of (a-c) real and (d-f) imaginary eigenfrequency surfaces in complex momentum space. Red and green stars mark the exceptional points. An increase in the dissipation rate lowers the imaginary eigenfrequency surfaces preferentially in the region of complex momenta with real part approaching the band edge. In each of the plots, two exceptional points are marked by red and green stars.

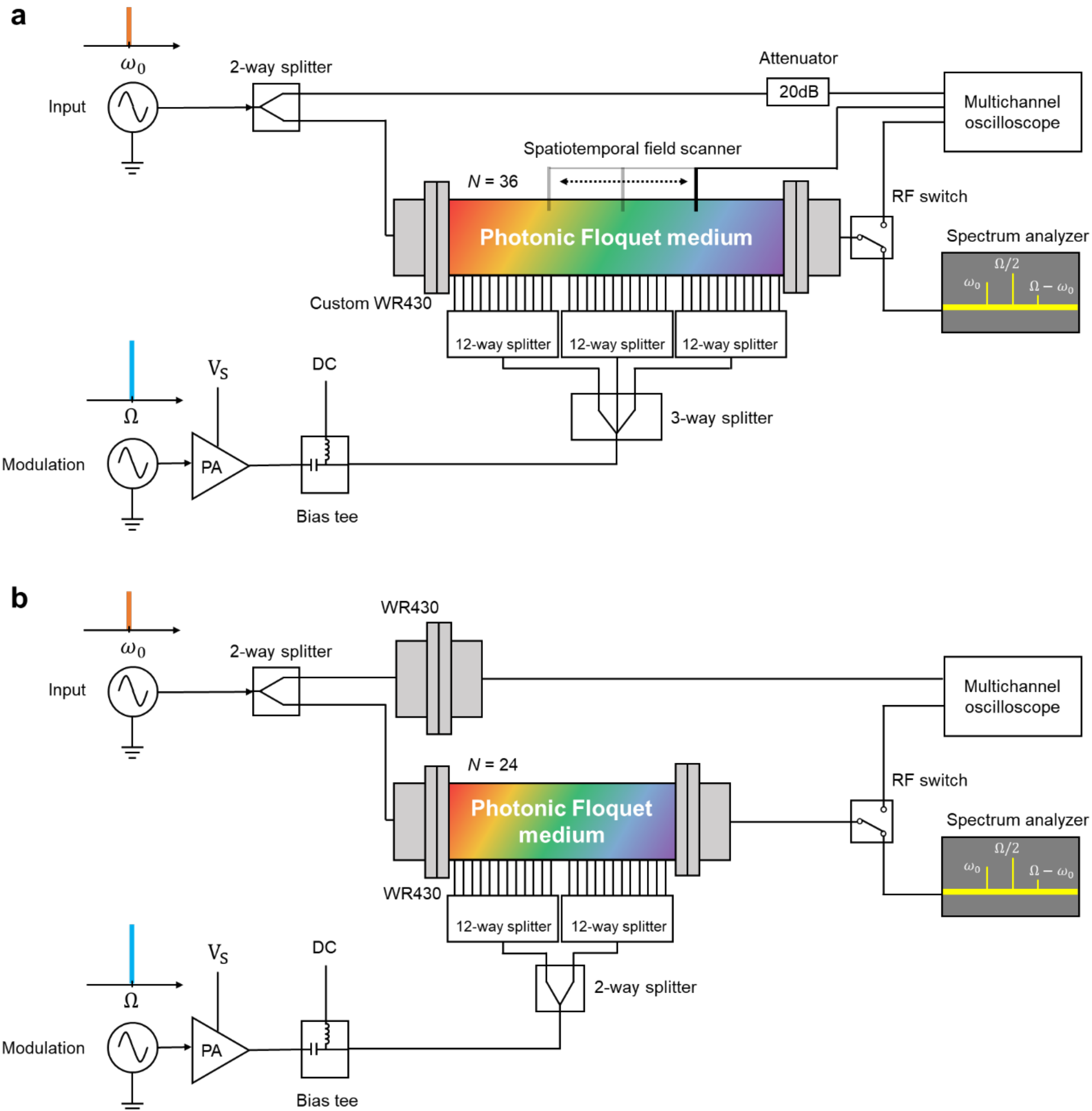

**Supplementary Figure S8. Schematics of the experimental setups.** (a) Schematic diagrams of the spatiotemporal field probing measurement setup and (b) phase retardation measurement setup. Here, the acronyms are as follows. PA: power amplifier; $V_s$: supply voltage; $N$: number of unit cells; DC: direct current; WR430: rectangular waveguide with opening dimensions of $4.30 \times 2.15$ inches.